\documentclass[prc,tightenlines,secnum,preprint,groupedaddress]{revtex4-1}

\usepackage{pifont}
\usepackage{color}
\usepackage{subfigure}
\usepackage{upgreek}
\usepackage{bm, braket}
\usepackage{subfigure}
\usepackage{multirow}
\usepackage{rotating}
\usepackage{longtable}
\usepackage{appendix}
\usepackage[usenames,dvipsnames]{xcolor}
\definecolor{light-gray}{gray}{0.6}
\usepackage{amssymb}
\usepackage{amsmath}
\usepackage{mathrsfs}
\usepackage{multirow}
\usepackage{lineno}

\newcommand{\molion}{$^3$HeT$^+$}
\newcommand{\hethree}{$^3$He}
\newcommand{\molT}{T$_2$}
\newcommand{\hhion}{$^3$HeH$^+$}
\newcommand{\etal}{{\em et al.}}

\newcommand*\patchAmsMathEnvironmentForLineno[1]{%
  \expandafter\let\csname old#1\expandafter\endcsname\csname #1\endcsname
  \expandafter\let\csname oldend#1\expandafter\endcsname\csname end#1\endcsname
  \renewenvironment{#1}%
     {\linenomath\csname old#1\endcsname}%
     {\csname oldend#1\endcsname\endlinenomath}}% 
\newcommand*\patchBothAmsMathEnvironmentsForLineno[1]{%
  \patchAmsMathEnvironmentForLineno{#1}%
  \patchAmsMathEnvironmentForLineno{#1*}}%
\AtBeginDocument{%
\patchBothAmsMathEnvironmentsForLineno{equation}%
\patchBothAmsMathEnvironmentsForLineno{align}%
\patchBothAmsMathEnvironmentsForLineno{flalign}%
\patchBothAmsMathEnvironmentsForLineno{alignat}%
\patchBothAmsMathEnvironmentsForLineno{gather}%
\patchBothAmsMathEnvironmentsForLineno{multline}%
}

\usepackage[pagebackref=false]{hyperref}

\hypersetup{
 colorlinks=true,     % color the words instead of use a colored box
 urlcolor=blue,       % \href{...}{...} external (URL)
 linkcolor=blue,     % \ref{...} and \pageref{...}
 citecolor=blue,     % \cite{}
 letterpaper=true,
 plainpages=false,
 breaklinks=true,
 bookmarksnumbered=true,
 bookmarksopen=true,
 pdfkeywords={},
 pdfpagemode=UseOutlines  % None, UseThumbs, UseOutlines, FullScreen
 }

\begin{document}

\title{Assessment of molecular effects on neutrino mass measurements from tritium beta decay}

\author{L.\,I.~Bodine}
\email[corresponding author: ]{lbodine@uw.edu}
\author{D.\,S.~Parno}
\email[]{dparno@uw.edu}
\author{R.\,G.\,H.~Robertson}
\email[]{rghr@uw.edu}
\affiliation{Center for Experimental Nuclear Physics and Astrophysics, and Department of Physics, University of Washington, Seattle WA 98195, USA}

\begin{abstract}
The beta decay of molecular tritium currently provides the highest sensitivity in laboratory-based neutrino mass measurements.  The upcoming Karlsruhe Tritium Neutrino  (KATRIN) experiment will improve the sensitivity to 0.2~eV, making a percent-level quantitative understanding of molecular effects essential.   The modern theoretical calculations available for neutrino-mass experiments agree with spectroscopic data.  Moreover, when neutrino-mass experiments performed in the 1980s with gaseous tritium are re-evaluated using these modern calculations, the extracted neutrino mass-squared values are consistent with zero instead of being significantly negative.  On the other hand, the calculated molecular final-state branching ratios are in tension with dissociation experiments performed in the 1950s.  We re-examine the theory of the final-state spectrum of molecular tritium decay and its effect on the determination of the neutrino mass, with an emphasis on the role of the vibrational- and rotational-state distribution in the ground electronic state.   General features can be reproduced quantitatively from considerations of kinematics and zero-point motion.   We summarize the status of validation efforts and suggest means for resolving the apparent discrepancy in dissociation rates. 
\end{abstract}


\maketitle

\tableofcontents

\section{Introduction}
\label{sec:intro}

The fact that neutrinos have mass~\cite{Fukuda:1998mi,Ahmad:2002jz} is the first definitive disagreement with the minimal Standard Model of particle physics.  As new extensions to the model are developed, a determination of the absolute neutrino mass scale will be essential~\cite{King:2014nza}. In addition, this mass scale influences the large-scale structure of the universe and is an important ingredient in cosmological models~\cite{Lesgourgues:2014zoa,PhysRevLett.112.051303}. Observables related to the neutrino mass are accessible through cosmological studies, neutrinoless double beta decay, and supernova neutrino observations. However, the most direct approach to the neutrino mass, with minimal model dependence, is by detailed measurement of the shape of the nuclear beta-decay spectrum near the endpoint.  

Tritium (T) undergoes an allowed nuclear beta decay, transforming to \hethree\  with the emission of a beta electron and electron antineutrino. The low Q-value of 18.6~keV means that the modification of the spectral shape by the neutrino mass is relatively large.  In addition the half-life of 12.3~years allows sources with high specific activity to be constructed.  

\begin{figure}[h]
\includegraphics[width=0.48\textwidth]{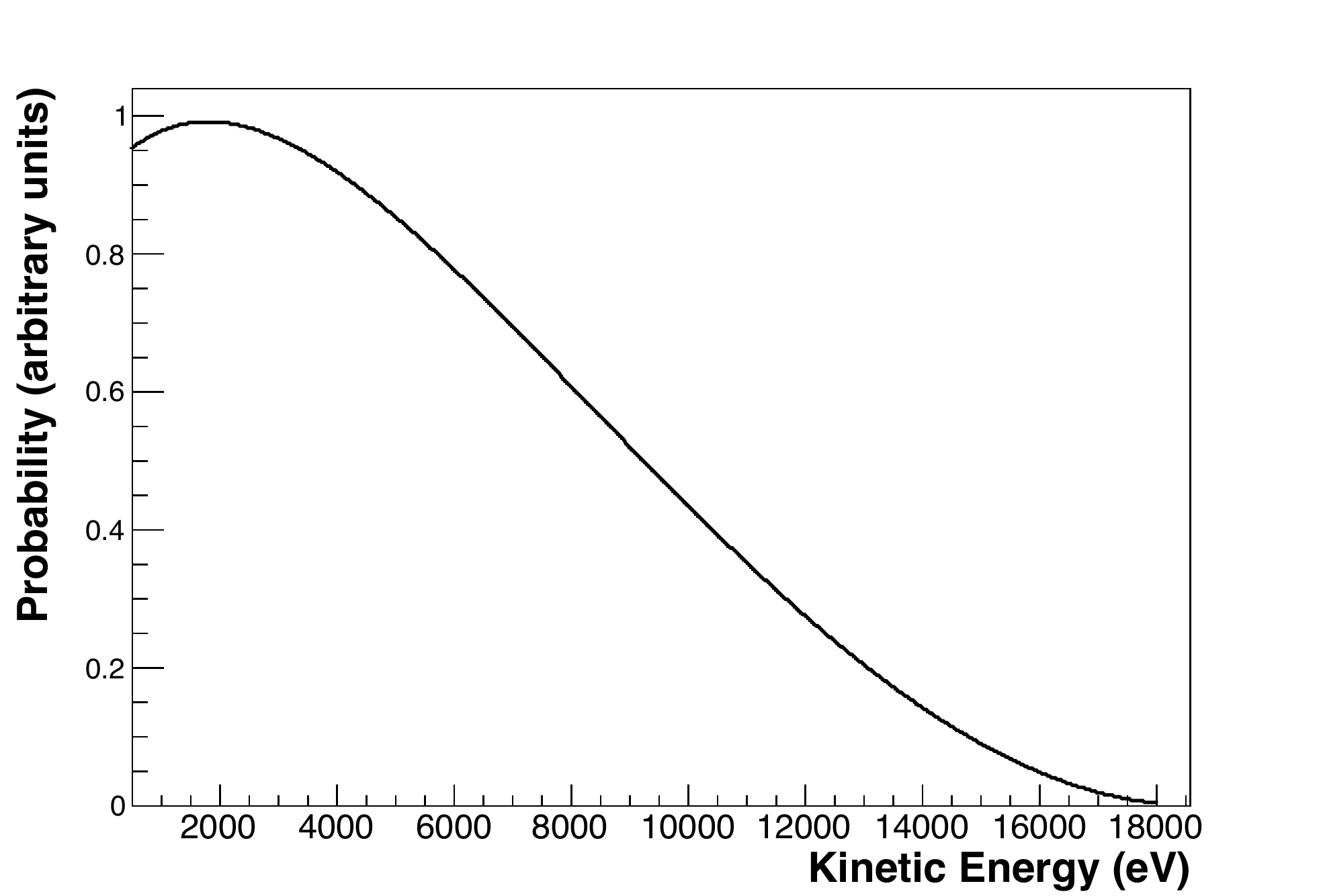}
\includegraphics[width=0.48\textwidth]{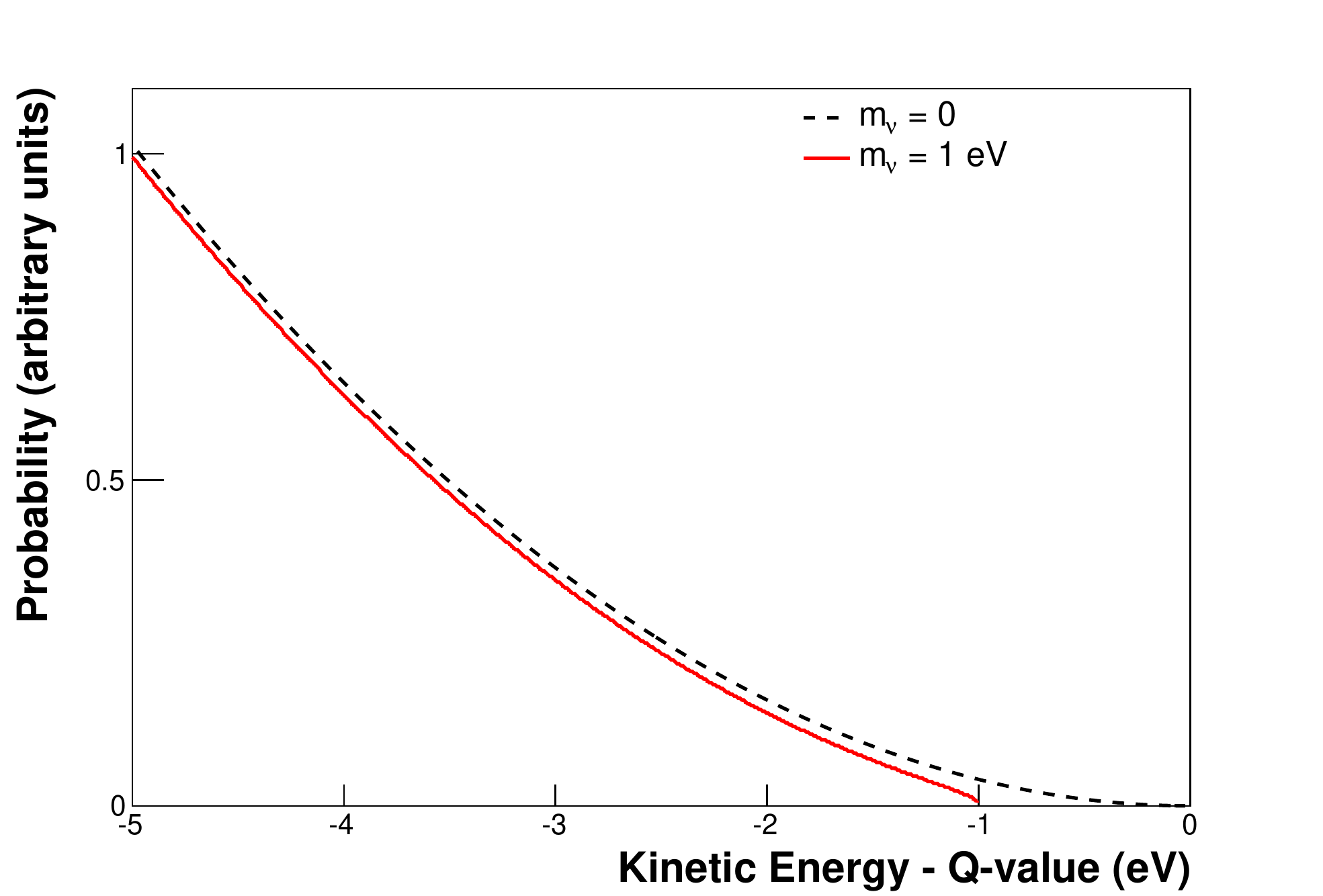}
 \caption{Tritium beta spectrum with 3 active neutrinos with masses $m_{\nu i} \simeq 1$ eV  for the case of no daughter excitation. The left panel shows the full spectrum.  The right panel shows the last 5~eV before the endpoint, with the dotted curve indicating the spectral shape for $m_{\nu i} = 0$.}
 \label{fig:numass}
 \end{figure}

The well-known form of the tritium beta spectrum is illustrated schematically in Fig.~\ref{fig:numass}  for massless neutrinos and for 1-eV neutrinos.  
It is the task of the experimentalist to measure the spectral shape and thereby determine the neutrino mass.   Only a fraction of order $10^{-13}$ of the decays populate the last 1~eV of the beta spectrum.  Uncertainty on the Q-value and practical experimental challenges preclude fixing the endpoint energy during data analysis and it is therefore treated as a fitted `nuisance parameter.'   Furthermore, the spectral distortion due to the neutrino mass is small and distortions of similar size can arise from a number of theoretical corrections and from instrumental effects.   For a molecular tritium source, the largest modifications to the spectrum are caused by excitations of the daughter molecule formed in the decay, which must be calculated from theory.  One could consider using a non-molecular source, such as T$^+$ or T, but these are far less practical due to space-charge limitations and the high reactivity of atomic hydrogen.

The ongoing construction of the Karlsruhe Tritium Neutrino experiment (KATRIN)~\cite{katrin04}, the next-generation tritium-based neutrino-mass experiment, has renewed interest in the molecular final-state distribution (FSD) populated by \molT{} beta decay~\cite{ott08}.  With a design neutrino-mass sensitivity of 0.2~eV, KATRIN depends critically on a theoretical understanding of molecular effects. Accordingly, extremely precise, {\em ab initio} calculations of the molecular final-state spectrum have been performed~\cite{saenz00,doss06} in the region of interest for KATRIN, near the endpoint of the beta electron energy spectrum. A direct experimental verification of these calculations through a study of the molecular final-state spectrum itself is not practical as explained in Sec.~\ref{sec:tests}.  Indirect tests can be performed, but have yielded mixed results. Although most of the spectrum of the HeH$^+$ isotopolog is inaccessible to experiment, many predicted spectral features have been observed in emission; HeH$^+$ photodissociation measurements are also compatible with theory, although a high-precision test has yet to be performed. On the other hand, measurements of the branching ratio of \molT{} to the bound molecular ion \molion{} following beta decay -- another observable indirectly related to the final-state distribution -- show stark disagreement with predictions. 

In this work, we discuss the aspects of the neutrino-mass measurements that have motivated study of the molecular final states excited in \molT{} beta decay  and summarize the current state of theoretical work on the topic. We  begin by examining the commonly used theoretical expression for the spectrum of allowed beta decay and derive a more general expression that facilitates a consistent treatment of molecular `final-state' excitations.    Focusing on the region of the spectrum near the endpoint, we show that the energy spread caused by molecular excitations is dominated by the  zero-point motion of the parent \molT{} molecule, and derive a general analytic expression for the variance of the ground-state manifold of states that includes not only zero-point vibration, but rotational and translational degrees of freedom.  The expression can be applied at any selected temperature up to 300 K,  to the 3 isotopologs \molT , DT, and HT,  to any chosen ortho-para admixture, and to address possible uncertainty in the rotational-state temperature.   The variance of the final state distribution is found to be quite sensitive to whether rotational thermal equilibrium has been achieved in the source gas.  We then examine several indirect experimental approaches for validating theoretical calculations of the final-state distribution,  and review existing measurements.   When modern calculations are used to re-evaluate gaseous tritium experiments performed in the 1980s, it is found that negative values of $m_\nu^2$ are eliminated.  We suggest desiderata for a new experimental investigation of the branching ratio to the ground-state manifold with a view to resolving the discrepancies of more than 50 years' standing. 

 \section{Direct neutrino mass measurements: experimental progress}
 \label{sec:exp_progress}
 
\subsection{Historical tritium-based neutrino-mass experiments}
\label{sec:t2_numass_history}

Tritium-based experiments to measure the absolute mass of the neutrino have a long history. Robertson and Knapp~\cite{robertson:1988aa} review early experiments, Otten and Weinheimer~\cite{ott08} give a detailed treatment of more recent experiments, and Drexlin \etal\ \cite{drexlin:2013} review experiments that are currently under construction.  

The issue of atomic and molecular excitations in tritium-based neutrino experiments was first raised by Bergkvist in the early 1970s~\cite{bergkvist:1971}.  He was able to set a 55-eV limit~\cite{bergkvist:1972aa} and noted that an understanding of daughter excitations was required to  improve limits further.   His work motivated the construction of an experiment with a windowless, gaseous \molT{} source at Los Alamos National Laboratory (LANL)~\cite{wilkerson:1987aa,robertson:1991aa}.  The use of \molT{} is advantageous  because the molecular final-state calculations are more tractable than for more complex sources, and a gaseous source minimizes the effects of scattering on the beta spectrum.   The LANL experiment yielded an upper limit of $m_\nu< 9.3$~eV at the 95\% confidence level~\cite{robertson:1991aa} with a 2-$\sigma$ excess of events observed in the endpoint region, reported quantitatively as a negative central value of $m_\nu^2$.  An experiment at Lawrence Livermore National Laboratory (LLNL), also using a windowless, gaseous \molT{} source, yielded a central value in good agreement with the LANL result, but with much reduced statistical uncertainties.  The excess of events  near the endpoint then corresponded to $6\sigma$~\cite{stoeffl:1995aa}.  

Concurrent experiments in Beijing~\cite{beijing:1993}, Tokyo~\cite{tokyo:1991} and Zurich~\cite{zurich:1992} used complex tritium sources.  All of these experiments gave results that were consistent with zero neutrino mass but with central values in the unphysical negative-mass-squared region, which is symptomatic of an underestimated theoretical or experimental contribution to the resolution function.  Attempts to reduce such influences furthered interest in molecular-tritium experiments, where {\em ab initio} molecular calculations were possible, and inspired further theoretical work on the molecular final-state distribution in the late 1990s (Sec.~\ref{sec:theory}).

The Particle Data Group evaluation~\cite{pdg2014} of the present limit on the neutrino mass, $m_\nu < 2$~eV at an unstated confidence level, is derived from the Mainz~\cite{mainz:2005} and Troitsk~\cite{troitsk:2003,troitsk:2011} experiments, both of which employed a new type of spectrometer. In a magnetic-adiabatic-collimation-with-electrostatic (MAC-E) filter~\cite{beamson:1980aa}, the momenta of beta electrons rotate to a mostly longitudinal direction as the electrons pass from a region of large magnetic field to a region of magnetic-field minimum. The kinetic energy of the resulting broad electron beam is then analyzed with a longitudinal retarding potential.  

The Mainz source consisted of \molT{} films quench-condensed onto substrates of highly oriented pyrolytic graphite. Solid-state source effects, such as dewetting effects and local lattice relaxation after the decay of a bound tritium atom, required careful attention in the Mainz analysis. The final Mainz result was $m_\nu < 2.3$~eV at 95\% confidence~\cite{mainz:2005}.

The Troitsk experiment, like its predecessors at LANL and LLNL, used a windowless, gaseous tritium source. The gas density and source purity were monitored indirectly by a mass analyzer at the source and by count-rate measurements at a low retarding-potential setting. During later runs an electron gun mounted upstream of the source was used to monitor the column density.  The initial analysis of the data required the inclusion of a step function added to the spectral shape~\cite{troitsk:2003}, the so-called ``Troitsk anomaly.''  The final Troitsk result, based on a re-analysis of the subset of runs for which electron-gun source-column-density calibrations  were available, was $m_\nu < 2.05$~eV at 95\% confidence~\cite{troitsk:2011}.  No step anomaly was required in the re-analysis. 

\subsection{Future prospects for direct neutrino-mass experiments}
\label{sec:t2_numass_future}

As the sensitivity of T$_2$-based experiments improves, an accurate understanding of the role of molecular final states after beta decay becomes increasingly important.   The systematic uncertainty associated with final states has been a major motivator in the search for other experimental approaches to direct neutrino mass measurement.  The common alternative approach employs microcalorimeters with sources of rhenium (MARE~\cite{ferri:2014}) or holmium (HOLMES~\cite{galeazzi:2012}, ECHo~\cite{blaum:2013, gastaldo:2013}, and a LANL experiment~\cite{engle:2013}).  Microcalorimeters suffer from pile-up spectral distortions, requiring the construction of a large number (order of millions) of functionally identical calorimeters.  

Alternative measurement techniques using tritium sources are also being explored. An approach for coincidence detection of the beta electron and the $^3$He$^+$ ion from a source of trapped tritium atoms was proposed~\cite{jerkins:2010} but later shown to be infeasible~\cite{otten:2011, jerkins:2011}. The Project~8 collaboration is currently studying the feasibility of measuring beta electron energies by trapping and measuring their cyclotron radiation frequencies with microwave antennae~\cite{monreal:2009aa, Project8_WhitePaper:2013}.  In its planned use of a \molT{} source, Project~8 again requires knowledge of the molecular final states of the source, although the collaboration is also studying the possibility of building an atomic T source by magnetically trapping single atoms as well as emitted electrons. Substantial research and development are required before a full experimental design can be developed.

Molecular-tritium beta decay remains the major focus of experimental work on the direct measurement of the neutrino mass.  Scheduled to begin taking data in 2016, the KATRIN experiment will be the most sensitive neutrino mass experiment to date with a design-sensitivity of 0.2~eV at the 90\% confidence level~\cite{katrin04}.  To achieve this level of sensitivity, the total systematic uncertainty must be controlled to within a budget of approximately $\sigma_{\rm syst}(m_\nu^2) \sim 17 \times 10^{-3} {\rm \ eV}^2$.  

The molecular final-state distribution populated by \molT{} decay represents one of the larger potential sources of systematic error in KATRIN.  A 1\% uncertainty in the calculated width of the ground-state molecular rotation and vibration distribution would contribute $6 \times 10^{-3} {\rm \ eV}^2$ to the budget for $\sigma_{\rm syst}(m_\nu^2)$~\cite{katrin04}.

Other  sources of systematic uncertainty for KATRIN are more amenable to experimental control~\cite{katrin04}. An electron gun behind the $10^{11}$~Bq windowless, gaseous \molT{} source will allow calibration of the experimental transmission function and of the energy loss experienced by electrons traveling through the source. The retarding potential of the KATRIN MAC-E filter will be independently monitored by the refurbished spectrometer from the Mainz experiment~\cite{slezak:2013} and by a high-voltage divider with a demonstrated stability of $6.0 \times 10^{-7}$ per month~\cite{thummler:2009}. Fluctuations in the column density of the source, which affect the scattering probability for electrons exiting the source, will be limited to the 0.1\% level through control of the tritium injection rate, the pumping speed, and the vessel temperature; a temperature stability of $ 5 \times 10^{-5}$ per hour at 30~K has been demonstrated with a prototype system~\cite{grohmann:2013aa}. In addition to the primary component \molT{}, it is expected that the KATRIN source will also contain DT and, to a lesser extent, HT.  To achieve the desired stability of the column density and column activity, the isotopic purity of the source must be determinable to a relative precision of  $2\times 10^{-3}$~\cite{katrin04}, and to this end the composition of the source gas will be monitored via Raman scattering in the tritium recirculation loop that feeds the source~\cite{fischer:2011aa,schlosser:2013}.

Today, the beta decay of molecular tritium provides the most immediate path to improving the sensitivity to neutrino mass by direct, laboratory determination. Both the anticipated sensitivity of the KATRIN experiment now under construction and the development of new ideas motivate a careful evaluation of the \molion{}  states excited in tritium beta decay.
  
\section{Form of the beta spectrum}
\label{sec:form_beta_spectrum}

The tritium decay process is accurately described by the Fermi theory of beta decay~\cite{fermi:1934}.  Tritium and helium-3 are mirror nuclei, so the nuclear matrix element $M_{\rm nuc}$ is maximal.  The transition is allowed, and the spectrum is not significantly modified by a shape factor dependent on the kinetic energy of the electron.   Hence the shape of the beta decay spectrum is determined by the neutrino mass $m_{\nu}$; electron mass $m_e$; total electron  energy $E_e$; maximum  energy of  the electron, $E_{\rm max} =Q - E_{\rm rec}^{\rm kin}+m_e$; and the energies $V_k$ and probabilities $P_k$ associated with excitations of the daughter ion.  The recoil energy $E_{\rm rec}^{\rm kin}$ consists of  translational kinetic energy of the daughter ion.  Since the discovery of neutrino oscillations shows there are three different neutrino eigenmasses ($m_{\nu i}$), the full spectrum becomes an incoherent sum over individual spectra for mass index $i=1,2,3$, with intensities given by the squares of the neutrino mixing matrix elements ($U_{ei}$)~\cite{Caldwell:2001zz}.  The resulting distribution of the electron  energy $E_e$ is shown in Eq.~\ref{eqn:spect},  in which $G_F$ is the Fermi weak-coupling constant, $\theta_C$ is the Cabibbo angle, $F(Z,E_e)$ is the Fermi function correcting for the interaction between the electron and the nucleus, and $\Theta(E_{\rm max}-E_e-V_k-m_{\nu i})$ is a Heaviside step function ensuring energy conservation~\cite{robertson:1988aa}.  Units are chosen where $c=1$. 
 \begin{align}
  \label{eqn:spect}
 \frac{dN}{dE_e} = &\frac{G_F^2 m_e^5\cos^2\theta_C}{2\pi^3\hbar^7} |M_{\rm nuc}|^2 F(Z,E_e) p_eE_e\\
\nonumber &\times\sum_{i,k}|U_{ei}|^2 P_k(E_{\rm max}-E_e-V_k)\sqrt{(E_{\rm max}-E_e-V_k)^2-m_{\nu i}^2}\\
\nonumber &\times \Theta(E_{\rm max}-E_e-V_k-m_{\nu i})
 \end{align}
 A number of small corrections to this basic spectral form have been identified over the years and have been summarized by Wilkinson~\cite{wilkinson:1991zz}.  At the time of his work, the effects he enumerated were for the most part negligible, but as experimental precision has advanced, their significance has as well.  Radiative corrections are the most important and have subsequently been  re-examined~\cite{Gardner:2004ib}.  A comprehensive and fully relativistic treatment of weak magnetism and induced terms may be found in Ref.~\cite{Simkovic:2007yi}.

Formally, Eq.~\ref{eqn:spect} also contains  inaccuracies in its treatment of rotational and vibrational molecular excitations.    The mass of the nucleus is considered to be infinite in deriving the electron-neutrino phase space, and nuclear recoil is then treated separately in determining the molecular translation, rotation, and vibration in the final state.  Electronic excitations represent energy unavailable to the outgoing leptons, and the modification to the phase space is appropriately captured by the appearance of $V_k$ in expressions for the electron energy.  However, a correct treatment of  rotational and vibrational excitations becomes ambiguous inasmuch as the appropriate recoil mass is not defined.   
In addition, the center-of-mass frame invoked for the decay described by Eq.~\ref{eqn:spect} is not related in any simple way to the center of mass of an object more complex than an isolated atom.  In a molecule, the atoms are always in motion, a source of Doppler broadening for the observed electron.  These issues can be avoided by consideration in a relativistic formalism of the full three-body phase space populated in the decay.

Because of the momentum imparted by the leptons to the recoil nucleus, the phase space is three-body rather than two-body everywhere except at the endpoint.  While it is standard to neglect this effect, doing so introduces a small spectral distortion.  More importantly, the three-body form permits a  self-consistent treatment of recoil effects.   The spectrum endpoint is given without ambiguity for any molecular system by conservation of the four-momentum for the full system.  An exact three-body, fully relativistic calculation for the phase-space density has been given by Wu and Repko~\cite{PhysRevC.27.1754} (see also Masood \etal~\cite{Masood:2007rc} and Simkovic \etal~\cite{Simkovic:2007yi}):
\begin{eqnarray}
\label{eq:full_3body_dN_dE}
\frac{dN}{dE_e} &=& C F(Z,E_e)\frac{p_eE_e}{\epsilon^2}\left(1-\frac{E_e}{M}\right)\times \nonumber \\
& & \times \sum_{i}(\Delta_i-E_e)|U_{ei}|^2\left[(\Delta_i-E_e)^2-m_{\nu i}^2\epsilon^2\right]^{1/2}\Theta(E_{ei,{\rm max}}-E_e) 
\end{eqnarray}
with the following definitions:
\begin{eqnarray}
C&=&\frac{G_F^2 m_e^5\cos^2\theta_C}{2\pi^3\hbar^7} |M_{\rm nuc}|^2  \\
\Delta_i &=& \frac{1}{2M}(M^2-M_{(f)}^2 + m_e^2+m_{\nu i}^2) \\
E_{ei,{\rm max}}&=& \frac{1}{2M}(M^2-M_{(f)}^2 + m_e^2-m_{\nu i}^2-2m_{\nu i}M_{(f)}) \\
\epsilon&=& 1-\frac{2m_e}{M}+\frac{m_e^2}{M^2}. 
\end{eqnarray}
We have here generalized Wu and Repko's result by introducing multiple neutrino mass eigenstates $m_{\nu i}$.  The mass $M$ ($M_{(f)}$) is the mass of the initial (final) atom or molecule,  including  associated atomic electrons and any excitation energy that may be present.  The quantity $\Delta_i$, an experimentally useful fit parameter, is the `extrapolated endpoint energy' that is obtained when the neutrino mass in the term in square brackets in Eq.~\ref{eq:full_3body_dN_dE} is set to zero.    The quantity $E_{ei,{\rm max}}$ is the  maximum energy of the electron for each neutrino eigenmass~\cite{Masood:2005aj}.   The electron-neutrino correlation modifies the spectrum at recoil order ($\sim m_e/M$) \cite{Simkovic:2007yi} and is not included here.  

Both initial- and final-state excitations can now be introduced explicitly by indexing $M$ and $M_{(f)}$ to become $M_j$ and $M_{{(f)}k}$, respectively. 
 For each pair of initial and final states $jk$ there is a corresponding Q-value,
 \begin{eqnarray} 
 Q_{kj} &=& M_j - M_{{(f)}k} -m_e
 \end{eqnarray}
 which is the kinetic energy released in the transition in the absence of neutrino mass.   A special case is the atomic mass difference between the neutral atoms T (mass $M_0=A$) and $^3$He (mass $M_{(f)0}+m_e=A^\prime$) in their ground states, which we denote $Q_A$:
 \begin{eqnarray}
 Q_A&=& A-A^\prime.
 \end{eqnarray}
 This corresponds to the Q-value for bound-state beta decay from ground state to ground state, the kinetic energy being delivered entirely to the neutrino and recoil.  (The term ``Q-value'' without qualification is used inconsistently in the literature, sometimes meaning $Q_{00}$ and sometimes $Q_A$.  For the atomic case, those quantities differ by the single-ionization energy of He, 24.59 eV.)  
 
\begin{figure}[tb] 
   \centering
   \includegraphics[width=0.6\textwidth]{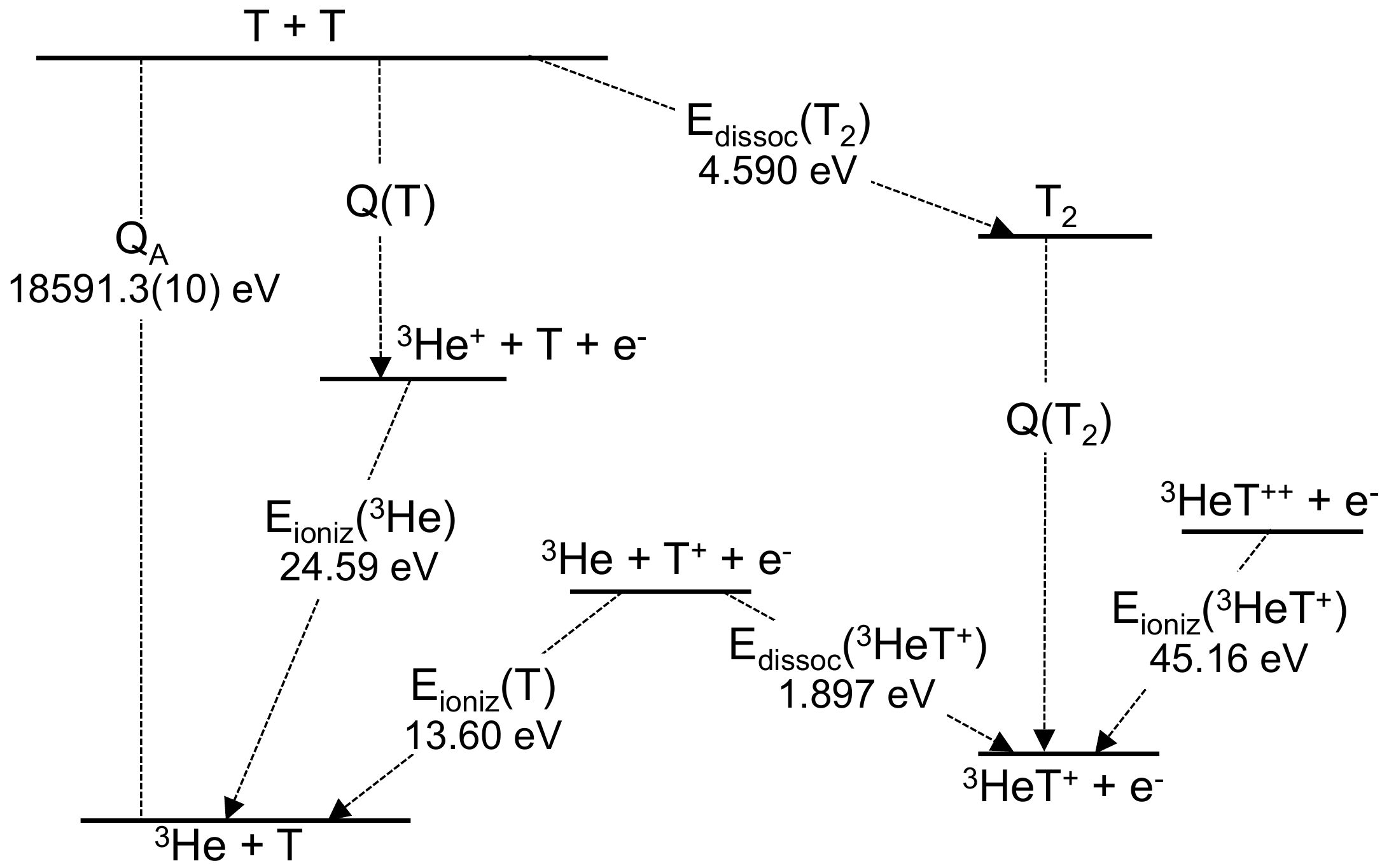} 
   \caption{Energy levels relevant to atomic and molecular tritium decay, patterned after Fig.~5 in Otten and Weinheimer~\cite{ott08}. The mass difference $Q_A$ is taken from Audi, Wapstra, and Thibault~\cite{audi:2003}. Dissociation energies are derived from calculations by Doss~\cite{dossphd:2007}; the ionization energy of \molion{} is from calculations by Ko\l{}os {\it et al.}~\cite{Kolos:1985}. The ionization energies for T~\cite{kramida:2010} and for $^3$He~\cite{morton:2006} are taken from recent compilations.}
   \label{fig:leveldiagram}
\end{figure}

In the general case the masses $M_j$ and $M_{(f)k}$ can be related to atomic masses by accounting for electron binding energies and for the possible presence of other atoms in the molecule: 
  \begin{eqnarray}
 M_j &=& A_s + A -b_j \\
 M_{{(f)}k} &=& A_s + A^\prime - b_{{(f)}k} -m_e \\
  Q_{kj} & = & Q_A -b_j + b_{{(f)}k}. 
  \end{eqnarray}
 Here, the binding energies $b_j$ and $b_{{(f)}k}$ are the energies released in transforming an atomic mass to the species of the parent or daughter, and the atomic mass of the other, `spectator,'  nucleus in the molecule (if present) is denoted $A_s$.  For example, the binding of two neutral tritium atoms to form a neutral \molT\ molecule in its ground state occurs with the release of $b_0 = +4.59$ eV.  Figure~\ref{fig:leveldiagram} is a graphical summary of the relevant binding energies.

  The extrapolated endpoint energy $\Delta_{ikj}$ can be expressed in terms of the corresponding Q-value: 
\begin{eqnarray}
\Delta_{ikj}&=& Q_{kj}+m_e - \frac{Q_{kj}}{2M_j}(Q_{kj}+2m_e)- \frac{m_{\nu i}^2}{2M_j}.
\end{eqnarray}
The extrapolated endpoint still has a dependence on neutrino mass, but it is completely negligible so the mass-eigenstate subscript $i$ on $\Delta$ will be omitted henceforth. The recoil-order term is small, a few parts in $10^4$ of  $Q_{kj}$.  Thus the extrapolated endpoint energy $\Delta_{kj}$ for excited final states ($\lesssim 100$ eV) can be taken to be the ground-state quantity $\Delta_{0j}$ minus the excitation energy.   

Weighting each transition by a matrix element $W_{kj}$ for the transition connecting the specific initial state $j$ to the final state $k$, the spectral density becomes 
\begin{eqnarray}
\left(\frac{dN}{dE_e}\right)_{kj} &=& C F(Z,E_e)\left|W_{kj}\right|^2\frac{p_eE_e}{\epsilon_j^2}(\Delta_{kj}-E_e)^2\left(1-\frac{E_e}{M_j}\right)\times \nonumber \\
& & \times \sum_{i}|U_{ei}|^2\left[(1-\frac{m_{\nu i}^2\epsilon_j^2}{(\Delta_{kj}-E_e)^2}\right]^{1/2}\Theta(E_{ei,{\rm max}(kj)}-E_e). \label{eq:betaspectrum}
\end{eqnarray}
 An expression for the matrix element $W_{kj}$ is given in Eq.~\ref{eqn:overlap} in Sec.~\ref{sec:theory}.

The maximum kinetic energy $E_{{\rm rec, max}(kj)}^{\rm kin}$ imparted to the recoil atom or molecule  is the difference between the extrapolated endpoint energy and the available mass energy in the decay:
\begin{eqnarray}
E_{{\rm rec, max}(kj)}^{\rm kin} & = & \frac{Q_{kj}}{2M_j}(Q_{kj}+2m_e)
\end{eqnarray}
A correct evaluation of the recoil energy is important because, as will be shown, the variance of the final-state distribution in the ground electronic state is directly proportional to it.

Table~\ref{tab:qvalues} summarizes the values of these parameters for several parent species, evaluated using the atomic mass difference $Q_A=18591.3(10)$~eV given by Audi, Wapstra, and Thibault~\cite{audi:2003}. In ref.~\cite{0295-5075-74-3-404}  a more recent measurement and a discussion of the experimental status of $Q_A$ are presented.

 \begin{table}[hbt]
\centering
%Caption above the table
\caption{Values in eV of the binding energies, Q-values, extrapolated endpoint energies,  and maximum recoil translational energies for five tritium-containing parents.  All of the quantities in the last three lines have the fractional uncertainty of $Q_A$.}
\label{tab:qvalues}
\vspace{0.1in}
\begin{tabular}{lrrrrr}
\hline 
\multirow{2}{*}{Quantity} 	& & &Parent \\ \cline{2-6}
& \phantom{aaaaaa} T$^+$  &\phantom{aaaaaa} T & \phantom{aaaaa} HT  & \phantom{aaaaa} DT  &\phantom{aaaaa} \molT   \\
\hline \hline
$b_0$ & -13.61&0&4.53&4.57&4.59 \\
$b_{(f)0}$ & -79.01&-24.59&-11.77&-11.73&-11.71 \\
$Q_{00}$ & 18525.85&18566.66&18574.96&18574.95&18574.95 \\
$\Delta_{00}-m_e$ & 18522.44&18563.25&18572.40&18572.91&18573.24 \\
$E_{{\rm rec, max}(00)}^{\rm kin}$ & 3.402&3.409&2.557&2.045&1.705 \\
\hline
\end{tabular}
\end{table}
In particular, it may be seen from the table that the endpoint energy for HT falls about 0.8 eV below that for \molT{}, and the endpoint energy for DT is intermediate between the two. However, the same underlying kinematics produce a compensating energy shift in the final-state distribution, as described in Sec.~\ref{sec:conceptualmodel}.

\section{Theory of molecular tritium beta decay}
\label{sec:theory}

Molecular states are specified by electronic ($n$), vibrational ($v$), rotational ($J$), and  azimuthal ($M$) quantum numbers.   For homonuclear molecules such as \molT\ the total nuclear spin ($I$) is important in satisfying the Pauli exclusion principle.  The \molT{} nuclear spin can take on two values, 1 and 0; $I=1$ corresponds to the triplet ortho state, and $I=0$ corresponds to the singlet para state.  The relevance of ortho and para states to the rotational quantum number and true molecular ground state is discussed in detail in Sec.~\ref{sec:isotopologs}.  

The final states excited in molecular beta decay include translational, electronic, rotational and vibrational excitations.  For the beta decay of an isolated tritium ion, only translational recoil is possible.  For a neutral tritium atom, precisely calculable electronic excitations also occur.  For a tritium molecule, rotational and vibrational excitations come into play and a theoretical treatment requires extensive computation.  Even for a parent molecule as simple as \molT{}, the electronic excited states of the daughter \molion{} molecule are complicated and unbound. Experimental advances, however, allow an important simplification: high statistics and excellent energy resolution will allow KATRIN to concentrate data taking within about 20~eV of the electron endpoint, a region in which electronic excitations play no role. Theoretical work can then focus on a precise calculation of the rotational and vibrational state distribution within the electronic ground state manifold. 

High-precision, {\em ab initio} calculations of the molecular excitations arising from \molT\ beta decay  have been performed~\cite{saenz00, doss06}.  The calculations use the Born-Oppenheimer approximation to factorize the molecular wavefunctions into electronic wavefunctions, vibrational wavefunctions and spherical harmonics dependent on the rotational and azimuthal quantum numbers.  Hyperfine structure is neglected except where spin symmetry must be respected in homonuclear systems.   Corrections to the Born-Oppenheimer and other approximations have also been investigated and found to be small~\cite{jon99}.  

\subsection{Geminal-basis method}
\label{sec:geminal_approach}

Theoretical investigations of beta decay in T$_2$ date back to the pioneering work of Cantwell in 1956~\cite{cantwell:1956aa}. Modern calculations are built on the theoretical framework of Ko\l{}os and Wolniewicz, who developed an adiabatic description of the hydrogen molecule in a basis of explicitly correlated two-electron wavefunctions in 1964~\cite{Kolos:1964}.  This basis is sometimes described as geminal because it treats the electrons as a pair rather than as independent particles.  Development of the geminal basis for the hydrogen molecule led to early calculations of the molecular effects in the decay of HT~\cite{Wolniewicz:1965}.  In a further refinement of the basis, Ko\l{}os \etal~\cite{Kolos:1985} investigated optimal parameter values.  The most recent calculations rely on those results with minor additional refinements~\cite{saenz00}.  

As Jonsell, Saenz, and Froelich~\cite{jon99} show, the transition matrix element related to the final-state \molion{} excitation $k\equiv (v_{(f)}, J_{(f)}, M_{(f)}, n_{(f)})$ from an initial \molT{} state $j\equiv (v, J, M, n)$ may be written,  

\begin{eqnarray}
\label{eqn:overlap}
\left|W_{kj}(K)\right|^2 &=&\left|\int\left[\chi_{v_{(f)} J_{(f)}M_{(f)}}^{n_{(f)}}(\bf R)\right]^*S_n(R)e^{i{\bf K\cdot R}}\xi_{vJM}^{n}  ({\bf R})d^3R\right|^2.
\end{eqnarray}

\noindent In this expression, $\chi$ and $\xi$ are the rotational-vibrational wave functions of the \molion \ and \molT \ molecules, respectively, and $S_n(R)$ is an electronic overlap integral.   The exponential of the dot product of the recoil momentum ${\bf K}$ and the nuclear separation ${\bf R}$ is a consequence of the recoil motion of the daughter He nucleus. 

The reduced mass of the daughter molecule enters into the radial Schr\"odinger equation, which must be solved in order to compute the rotational and vibrational energy levels. There is some ambiguity in the definition of this quantity, which depends on whether and how the masses of the two bound electrons are included. Coxon and Hajigeorgiu~\cite{coxon:1999}, comparing predicted energy levels to spectroscopic measurements (Sec.~\ref{sec:tests_energy_levels_HeH}), achieved the best agreement with an effective reduced mass that assumes one electron belongs strictly to the He nucleus, with the second electron distributed evenly between the H and He nuclei. Doss~\etal~\cite{doss06}, confirming this result, introduced the effective reduced mass to the calculation of the final-state distribution, but noted that the change was insignificant at the 0.1-eV level of foreseeable \molT{}-based neutrino-mass measurements.

\begin{figure}[tb] 
   \centering
    \includegraphics[width=0.8\textwidth]{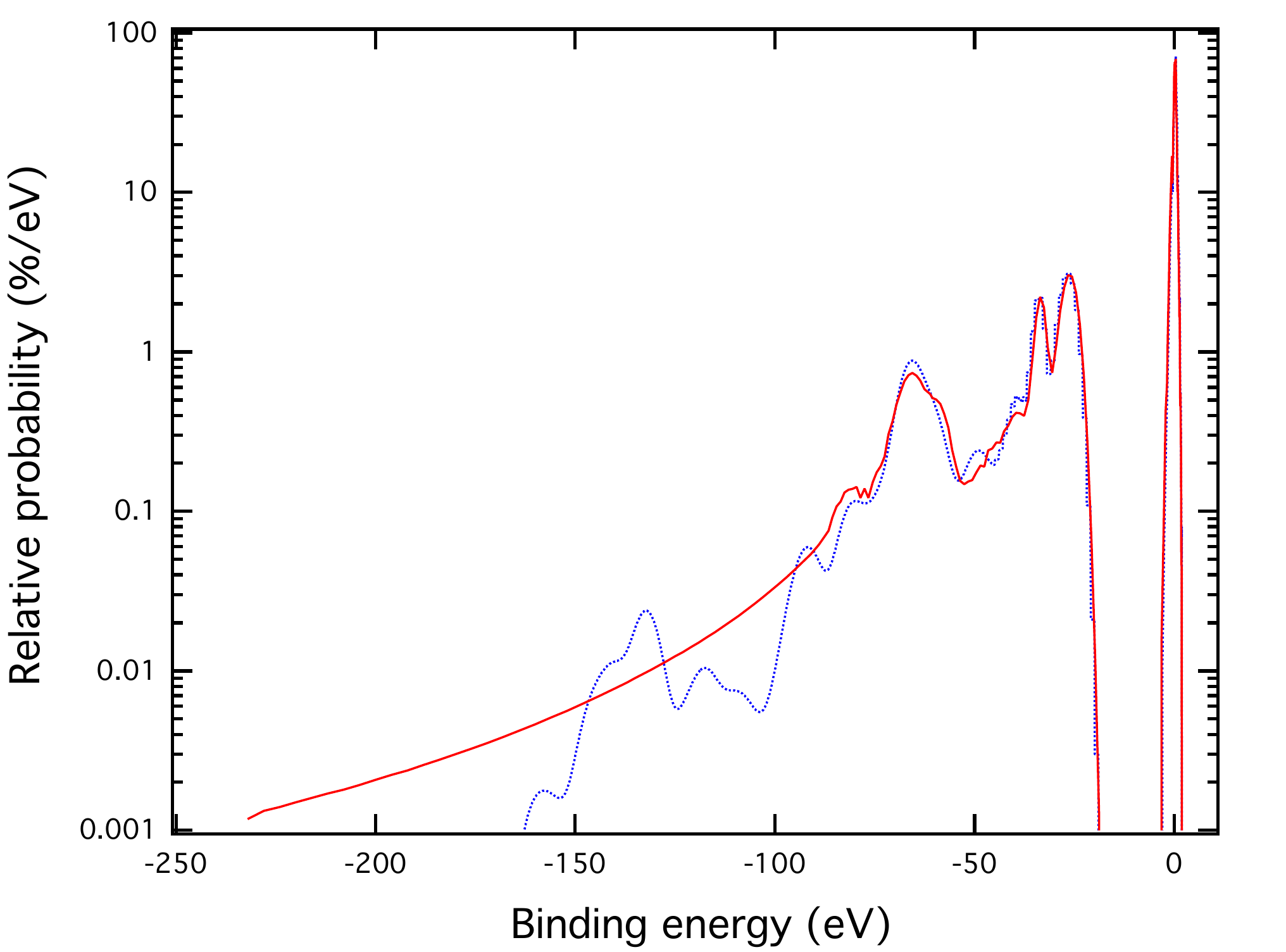} 
    \caption{Molecular spectrum excited in the beta decay of \molT\ ($J=0$) as calculated by Saenz {\em et al.}~\cite{saenz00} (solid curve, red online) and by Fackler {\em et al.}~\cite{fackler:1985} (dotted curve, blue online).  For the purposes of display and comparison, discrete states in the latter spectrum have been given a Gaussian profile with a standard deviation of 3 eV.}
   \label{fig:thspect}
\end{figure}

Fig.~\ref{fig:thspect} shows the spectrum of final-state molecular excitations from the beta decay of \molT\ ($J=0$ initial state) as published by Saenz {\em et al.}\ in 2000~\cite{saenz00}, compared with the 1985 calculation by Fackler {\em et al.}~\cite{fackler:1985}.  The electronic ground state appears as a large peak centered at 1.7~eV excitation energy (0.2 eV binding energy), broadened by the rotational and vibrational excitations.   The higher electronic states also suffer broadening as shown.  For a detailed description of the differences between the Saenz and Fackler calculations see reference~\cite{saenz00}.  The more recent results of Doss~\textit{et al.}~\cite{doss06} were not published in tabular form but a subset of the tables was provided to the KATRIN collaboration courtesy of Doss.  Reference~\cite{dossphd:2007} compares the Doss~\etal~\cite{doss06} results and the Saenz~\etal~\cite{saenz00} results. The differences are negligible for the ground state but noticeable in the electronic continuum, particularly above 45~eV of excitation energy (see Sec.~\ref{sec:continuum}).  

Unfortunately, in the geminal basis the convergence of the calculations depends on the explicit choice of basis functions and in-depth study has revealed that adding even one basis function can dramatically change the contributions of other functions~\cite{Bishop:1978}.  Significant optimization work was done to select the correct basis functions for \molT\ and \molion\ and provide reliable results~\cite{Kolos:1985}.  In lieu of explicitly computing uncertainties, which is impractical due to the volatility of the basis, published calculations typically include the deviation from 1 of the cumulative probability function at the highest excitation energy. However, this single number, while informative, is an insufficient gauge of accuracy. Despite the linear dependences and instability of the geminal basis, it has been used to produce essentially all final-state distribution calculations of the ground-state manifold since its publication~\cite{doss06,saenz00}.  

\subsection{Configuration-interaction method}
\label{sec:ci_approach}

The configuration-interaction (CI) method presents an alternative approach to modeling two-electron, diatomic molecules such as \molT{} and \molion{} within the Born-Oppenheimer approximation. In the CI technique, two-electron configurations are constructed as the products of pairs of solutions to the single-electron Schr\"odinger equation (denoted orbitals). Superpositions of these configurations are then used to build wavefunctions and make calculations. As the simplest two-electron heteronuclear molecule, HeH$^+$ was an early test bed for the method (see, {\it e.g.},~\cite{anex:1963,stuart:1964, michels:1966}). In the 1980s, parallel to the refinement of the geminal-basis method, the CI method was applied to the FSD following beta decay in \molT{}. Fackler {\it et al.}~\cite{fackler:1984} performed a preliminary study of decays to the first five electronic states of \molion{}; Martin and Cohen~\cite{martin:1985} used a more flexible basis set of Cartesian Gaussian orbitals to study the first 50~eV of the electronic continuum. (See Sec.~\ref{sec:continuum}.) Without the benefit of modern computation, however, such early treatments were neither complete nor precise enough to address the final-state spectrum in the region of interest for modern tritium-based neutrino-mass experiments such as KATRIN. 

More recently, Vanne and Saenz~\cite{vanne:2004} have developed a CI approach, based on an underlying B-spline basis set and carried out in an elliptical box, that shows promise for neutrino-mass experiments. This method avoids the linear dependences that tend to arise in numerical calculations with the geminal-basis method, allowing application to larger internuclear distances $R$ as well as the use of larger basis sets. Adding individual basis functions does not introduce artificial resonances. The discretization provided by the elliptical box allows the electronic continuum to be discretized as well, permitting the consideration of both bound and continuum states within the same basis set. Since all configurations are expressed in terms of one-electron wavefunctions, however, two-electron correlations are treated less accurately than in the geminal-basis method, especially if the configuration set is small.

Vanne and Saenz have compared their B-spline-based CI treatment of HeH$^+$ photoionization~\cite{vanne:2004} against one using the standard geminal basis~\cite{saenz:2003}. The first resonance in the $X^1\Sigma \rightarrow {^1\Sigma}$ photoionization cross section, at about 16~eV, is shifted about 0.5~eV higher in the CI results, likely due to the difference in treating two-electron correlations. The two approaches predict the same amplitude for this resonance and give good agreement for other features of the spectrum. 

The application of this method to tritium beta decay is a work in progress~\cite{Saenz:privcomm}. Once sufficiently complete configuration sets are calculated for \molT{} and for \molion{}, the electronic overlap integrals $S_n(R)$ can be computed. Transition probabilities may then be determined using Eq.~\ref{eqn:overlap}.

\subsection{Electronic continuum}
\label{sec:continuum}

The energy window for the KATRIN neutrino-mass measurement is narrow enough that related FSD calculations can focus on the \molion{} electronic ground state. However, it has been suggested that a measurement of the tritium beta spectrum over a wider energy range could be used to search for sterile neutrinos with mass on the eV scale~\cite{Formaggio:2011} or even on the keV scale~\cite{mertens:2014}. If the acquisition window extends more than about 40~eV below the beta endpoint, the analysis must account for the electronic continuum portion of the FSD. Table~\ref{tab:continuum} gives a brief overview of the variety of methods that have been applied to the problem. In addition to their differences in general approach, the available calculations differ in baseline assumptions. Early calculations often used the clamped-nuclei approximation rather than explicitly accounting for nuclear motion that broadens resonances. Assumptions about the localization of resonances can introduce errors at higher excitation energies~\cite{doss:2008}. Variation of the internuclear distance shifts the overall probability distribution but can also change the relative intensities of the electronic resonances~\cite{saenz:1997ii}. A significant simplification is possible at excitation energies above $\sim 200$~eV, a region in which the fast-moving ejected electron sees the $^3$He$^{++}$ ion as equivalent to a bare He nucleus. The high-excitation-energy tail of the FSD can then be described with a  spectrum adapted from the decay of atomic tritium~\cite{saenz:1997ii}. 

 \begin{table}[hbt]
\centering
%Caption above the table
\caption{Selected calculations of the probability $P_{\rm cont}$ of populating the electronic continuum of \molion{} in \molT{} beta decay. The integration range differs between the calculations, and the bounds are specified as excitation energies above the \molion{} ground state.}
\label{tab:continuum}
\begin{tabular}{llcc}
\hline
\textbf{Method} & \textbf{Reference} &  $\mathbf{P}_{\mathbf{cont}}$ & \textbf{Integration Range} \\
\hline
\hline
Complex scaling & Froelich {\it et al.} (1993)~\cite{froelich:1993} & $12.77\%$ & $45 - 90$~eV \\
Stieltjes imaging \phantom{aa} & Martin and Cohen (1985)~\cite{martin:1985} & $13.42\%$ & $45 - 94$~eV \\
Stabilization & Fackler {\it et al.} (1985)~\cite{fackler:1985} & $14.2\%$\phantom{0} & $45 - 200$~eV \\
$R$-matrix & Doss and Tennyson (2008)~\cite{doss:2008} \phantom{aa} & $13.66\%$ & ca.~$40-240$~eV* \\
\hline
\multicolumn{4}{l}{\footnotesize *Lower integration bound is not explicitly given.}
\end{tabular}
\end{table}

The calculated percentage of tritium decays that populate the electronic continuum is relatively consistent despite dramatic differences in the integration range, reflecting the fact that this region of the spectrum is dominated by a few autoionizing states near the ionization threshold. However, comparisons between different calculations, performed {\it e.g.} in Ref.~\cite{doss:2008} and~\cite{froelich:1993}, show significant discrepancies in the detailed structure of this part of the spectrum. For a sterile-neutrino search, knowledge of the integrated probability $P_{\rm cont}$ is not sufficient. If not properly accounted for, small structures in the FSD at high excitation energies could lead to errors in interpretation, especially when small mixing angles are considered. Sensitivity calculations for such a search must be guided by theoretical studies of this region of the FSD spectrum. 

\subsection{Molecular forms of tritium}
\label{sec:isotopologs}

The tritium-containing hydrogen isotopologs (HT, DT and \molT) have different reduced masses and thus different excitation spectra.  While the overall structure of the final-state spectrum remains qualitatively the same across isotopologs, the vibrational energy levels are  shifted and the probability of a transition to any specific rotational-vibrational state changes.  For example, the electronic excitations in \hhion\ are shifted $\sim$1~eV lower than the corresponding excitations in $^3$HeT$^+$~\cite{jon99}.      As shown in Table~\ref{tab:qvalues}, however, the difference in recoil mass also changes the extrapolated endpoint, canceling the change in the beta energy to first order~\cite{katrin04}.  

In addition to differences in reduced mass, nuclear spin and symmetry considerations play an important role in determining the allowed angular-momentum states of the homonuclear \molT\ molecule but do not apply to the heteronuclear DT and HT molecules.  In accordance with Fermi statistics, the overall \molT\ wavefunction must be antisymmetric under exchange of the tritium nuclei.   The electronic, rotational, and vibrational wavefunctions of the molecule are inherently symmetric.  Thus the spin-symmetric ortho state must be matched with an antisymmetric spatial wavefunction corresponding to odd $J$.  The spin-antisymmetric para state must be matched with a symmetric spatial wavefunction corresponding to even $J$.  Hence the ground state of the molecule is the para state with $J=0$. 

In thermal equilibrium the partition function of rotational states ($J$) in \molT{} may be written,
\begin{eqnarray}
\label{eqn:partition}
Z_{\rm equil} &=& \sum_{J=0}^{\infty}[2-(-1)^J](2J+1)e^{-J(J+1)\hbar^2/2\mathscr{I}k_BT},
\end{eqnarray}
to first order.  Here the first factor is the spin statistical weight for ortho (odd $J$) or para (even $J$) in the case of a homonuclear molecule, when total antisymmetry must be enforced, and $k_BT$ is the thermal energy.  The moment of inertia, $\mathscr{I}$, is related to the energy of the first excited state, $E_{J=1}$,
\begin{eqnarray}
\mathscr{I}&=& \frac{\hbar^2}{2E_{J=1}}.
\end{eqnarray} 
Since $E_{J=1} = 0.00497$~eV  \cite{dossphd:2007} is small compared to $k_BT$ at room temperature, the ortho-para ratio of a thermally equilibrated source at room temperature is essentially the ratio of the spin statistical weights, 3:1~\cite{jon99}.  Rather than the ortho-para ratio, the state of a molecular hydrogen source is typically characterized in terms of the parameter $\lambda$ quantifying the fraction of the source that is in the ortho state.   

The ortho-para transition requires a simultaneous change in the spin and rotational quantum numbers, making the ortho state metastable.   Thus transitions to lower rotational states are dominated by intrinsically slow quadrupole transitions.  For this reason, unless specific steps are taken to ensure it, thermal equilibrium of the rotational states of \molT{} cannot be guaranteed.  Thermalization of the spin degrees of freedom in a homonuclear hydrogen source is a slow, exothermic process, and uncertainty arises from the use of sources that are not in thermal equilibrium and that contain a mixture of states.   

Previous studies of molecular hydrogen have focused on the ortho-para ratio alone as the determining factor in the rotational-state distribution, a reasonable assumption for light isotopologs.  However, for \molT{} above cryogenic temperatures, states higher than $J=1$ have significant populations and the evolution of the full rotational-state distribution must be considered.  Spontaneous quadrupole transitions are extremely slow, on the order of $10^{-7}$~s$^{-1}$  in free space~\cite{wol:1998}, and transitions will be dominated by collisions with other tritium molecules and the walls.   The rate of these processes depends on the detailed design of the gas system and must be carefully modeled to determine the rotational-state distribution of the source.  

\section{Conceptual model of the rotational-vibrational spectrum}
\label{sec:conceptualmodel}

As we have seen in Sec.~\ref{sec:theory}, a precise treatment of the molecular final-state spectrum requires an extensive theoretical framework. However, as experimental sensitivity has advanced, dependence on the highly excited states has diminished.  The width of the ground-state manifold now sets the fundamental limit on the sensitivity of experiments using \molT.  With the intention of gaining some insight into the physical origin of the width of this manifold we have developed a  simplified treatment, based on kinematic considerations and the approximation of the molecule as a simple harmonic oscillator.  It reproduces several features of the precisely calculated spectrum while clarifying the underlying physics. 

Qualitatively, the beta spectrum is influenced in two distinct ways by the molecular structure.  The rotational, vibrational and translational motions of the parent \molT{} molecule lead to modulation of the energy of the detected beta electron.  Some motions are essentially thermal in origin and contribute a Doppler shift in the laboratory electron energy.   Classically, each degree of freedom contains on average $\frac{1}{2}k_BT$ of energy, and the atomic velocity adds vectorially to the electron velocity.  Nevertheless, as we shall see, it is a uniquely quantum-mechanical effect, zero-point motion, that in fact dominates the spectrum at low temperatures.

In the following, our interest is in the rotational and vibrational degrees of freedom in the electronic ground state.   We begin by examining the purely kinematic constraints on the recoil momentum ${\bf p} = \hbar {\bf K}$.  We then, in a semiclassical approach, combine the initial momentum of the decaying T nucleus in the parent molecule with the momentum delivered by lepton recoil in order to find the momentum spectrum of the daughter \hethree{}.  Applying kinematic constraints, the momentum spectrum is expressed in terms of the corresponding translational and  excitation energies of the recoil molecular ion \molion \ or \hhion{}, for the parents \molT \ and HT, respectively.  

\subsection{Recoil momentum}
\label{sec:conceptual_recoilmomentum}

 The three-momentum imparted to the molecular system by the beta decay has a magnitude
  \begin{eqnarray}
 p&=& |{\bf p_e + p_\nu}| \nonumber \\
p^2 &=& E_e^2-m_e^2+(E_{\rm max}-E_e-E_{\rm rec}^{\rm kin})^2-m_\nu^2+2E_e(E_{\rm max}-E_e-E_{\rm rec}^{\rm kin})\beta \beta_\nu \cos\theta_{e\nu} \label{eq:recoilenergy}
 \end{eqnarray}
where $\theta_{e\nu}$ is the angle between the electron and the neutrino momenta, and $\beta$ and $\beta_\nu$ are, respectively, the electron speed and neutrino speed relative to the speed of light.  It is sufficient for the present purpose to neglect neutrino mass and also the kinetic energy of the recoil $E_{\rm rec}^{\rm kin}$ as it contributes corrections  of order $m_e/M \simeq 10^{-4}$ to the square of the recoil momentum.    

The electron-neutrino correlation term may be written \cite{Jackson:1957zz} 
\begin{eqnarray}
\left[1+a_{e\nu}\frac{{\bf p_e \centerdot p_\nu}}{E_eE_\nu}\right] = 1+a_{e\nu}\beta\cos{\theta_{e\nu}}.
\end{eqnarray}

\begin{figure}[tb]
 \includegraphics[width=0.5\textwidth]{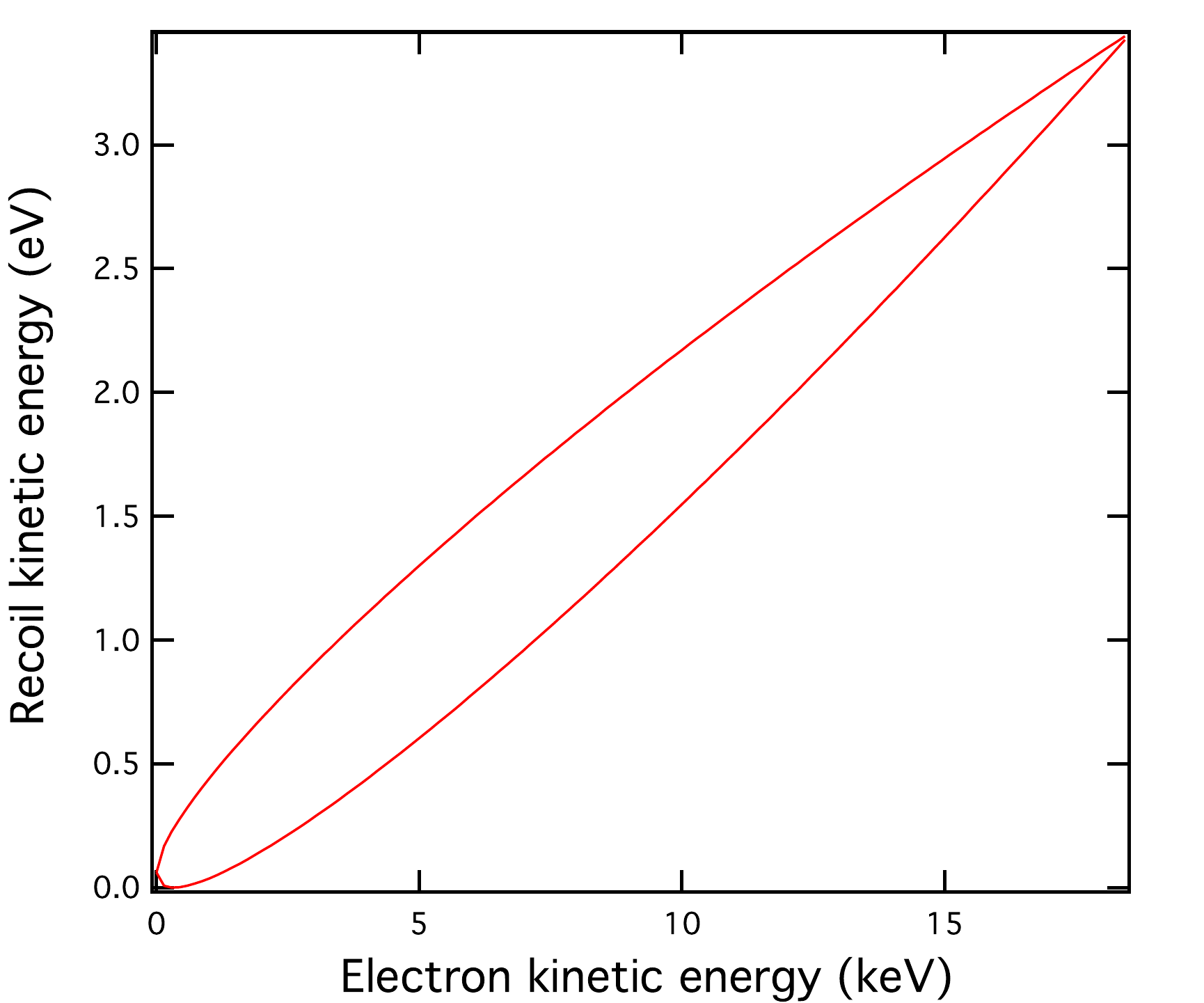}
 \caption{Recoil kinetic energy imparted to a \hethree \  daughter by the beta decay of an isolated tritium nucleus at rest.  The upper boundary of the envelope corresponds to $\theta_{e\nu}=0$ and the lower one to $\theta_{e\nu}=\pi$.}
 \label{fig:recoilenergy}
 \end{figure}

\noindent Using for $a_{e\nu}$ the value measured for the free neutron, $a_{e\nu}=0.105(6)$~\cite{Byrne:2002tx}, and noting that the electron velocity $\beta  \leq 0.26$, one sees that the electron-neutrino correlation is very weak in tritium decay.   The recoil-energy envelope for the decay of an isolated tritium nucleus is shown in Fig.~\ref{fig:recoilenergy}.

Although the recoil momentum is given immediately from the lepton momentum via momentum conservation, determining the recoil {\em energy} requires knowledge of the recoil mass.  In the case of an isolated T atom, shown in Fig.~\ref{fig:recoilenergy}, the calculation is unambiguous, but for a \molT \ molecule it is not.  For a very tightly bound system with no accessible internal degrees of freedom the mass would be the total mass (6 u), and for a very weakly bound one it would be 3 u.  Without further information, the recoil energy can be bounded above and below by kinematics and at these limits is entirely translational kinetic energy.  At the endpoint of the beta spectrum, 
\begin{eqnarray}
1.705 &\leq& E_{\rm rec}^{\rm kin} \leq 3.410 {\rm \ eV}.
\end{eqnarray}

The \molion \ ion has a spectrum of rotational and vibrational excitations that are one or two orders of magnitude smaller than the recoil energy, less like the strongly bound picture and more like the weakly bound one.   Some insight into the behavior of this system can be gained by considering first a purely classical \molT \ molecule at 0 K, such that both atoms are bound together but at rest.    If the molecule remains bound after beta decay, conservation of linear momentum requires that 1.705 eV must be in the form of translational kinetic energy, leaving only 1.705~eV available for internal excitations.  The binding energy of the final-state molecular ion \molion \ is 1.897 eV~\cite{dossphd:2007}, and, since this is greater than the available excitation energy, the \molion \ must remain bound in this classical picture with no thermal motion.     Then  the final state consists of a mass-6 ion with a translational kinetic energy of 1.705~eV and rotational and vibrational excitations totaling 1.705~eV.   How the excitation energy is apportioned between rotational and vibrational excitations depends (classically) on the relative orientation of the axis connecting the atoms to the lepton momentum direction, but the total excitation energy is always 1.705~eV. 

\begin{figure}[tb]
\includegraphics[width=7.8cm]{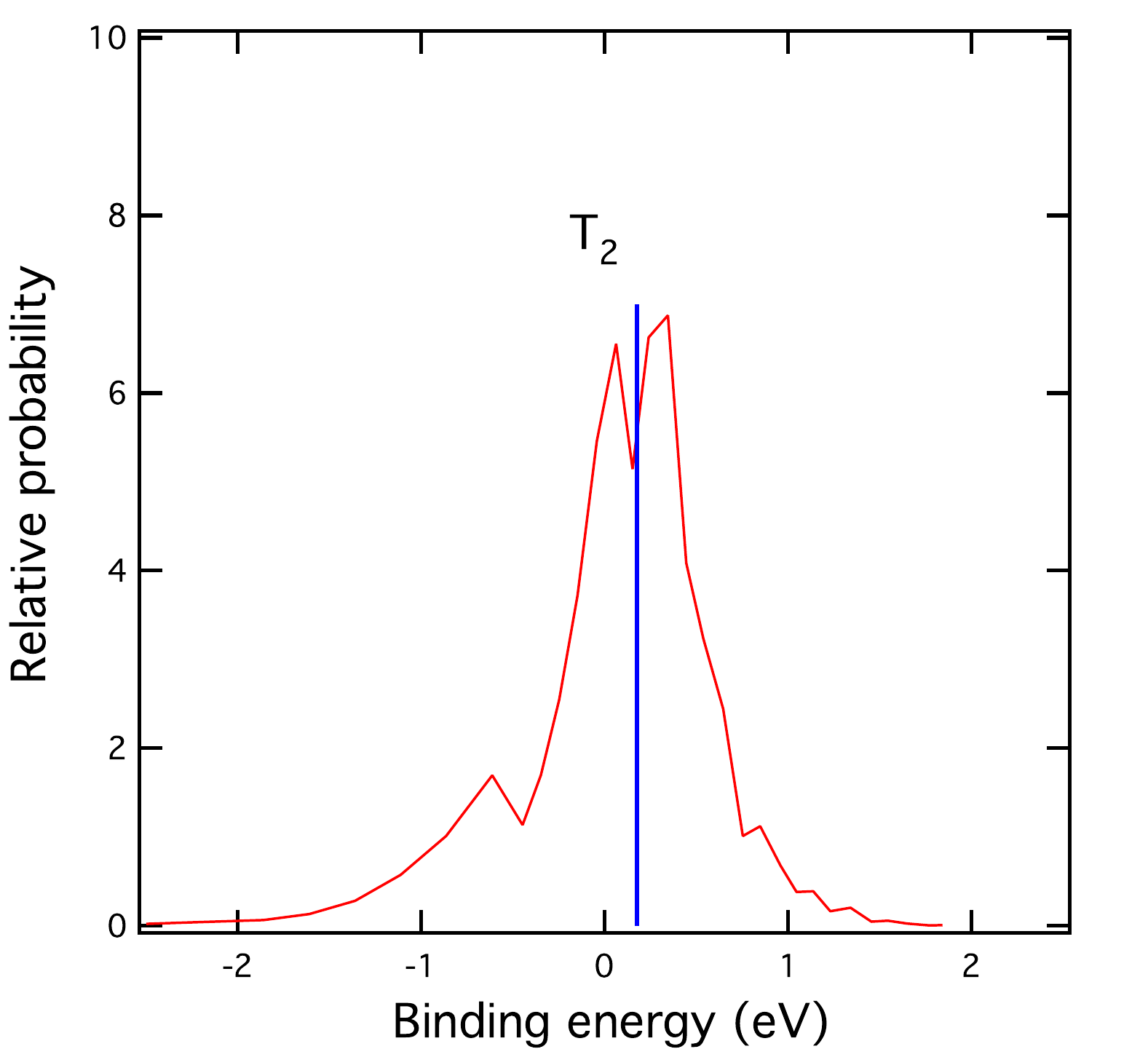}
\includegraphics[width=7.8cm]{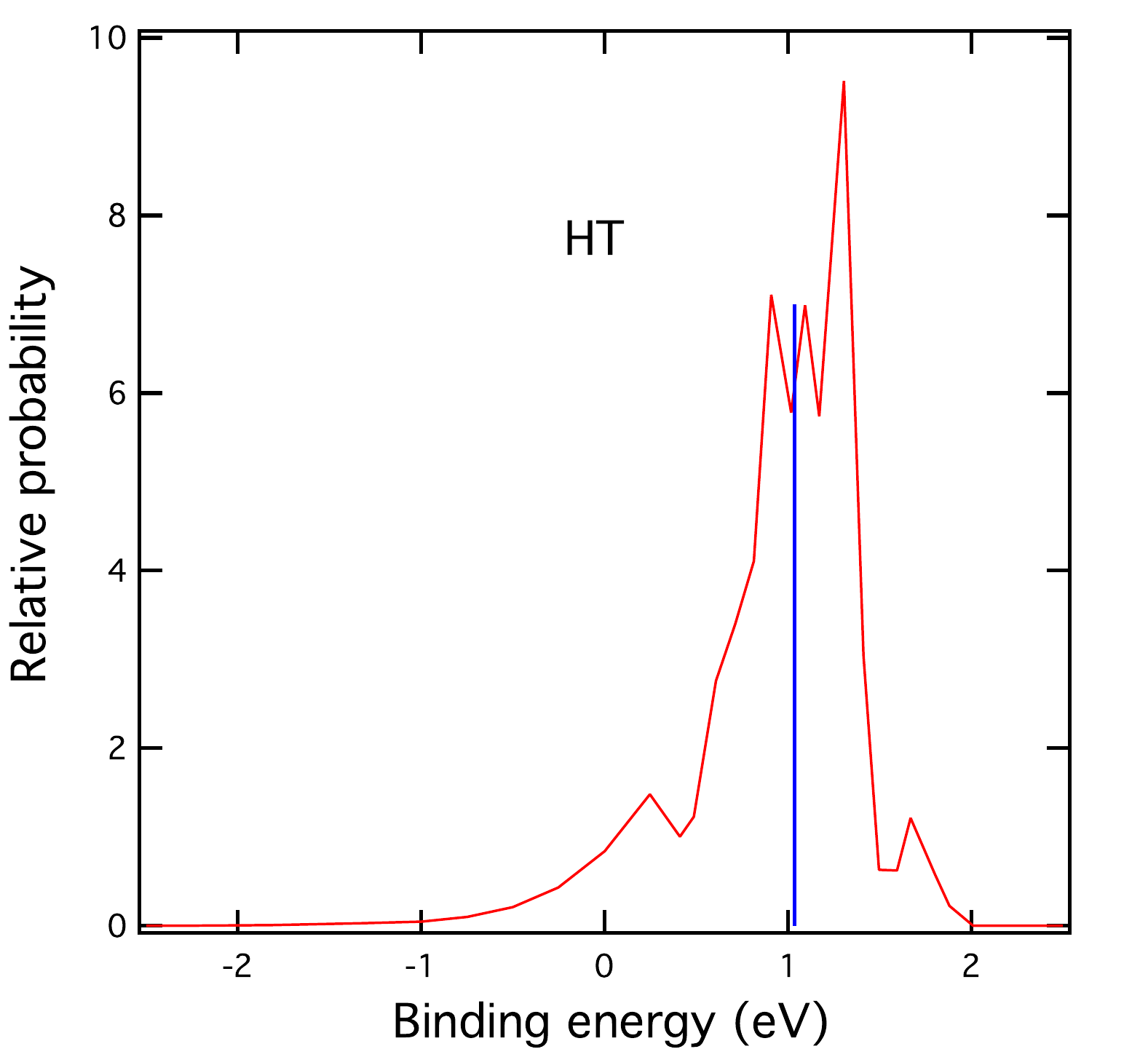}
 \caption{Distributions of excitation energy in the ground-state rotational and vibrational manifold of \molion \ (left) and \hhion \ (right) as calculated by Saenz {\em et al.} \cite{saenz00}. The expected value for the excitation energy in each case, based on kinematic considerations, is indicated by a vertical line.  An excitation energy of 0 corresponds to a binding energy of 1.897 eV~\cite{dossphd:2007}.}
 \label{fig:saenzcompare}
 \end{figure}

The equivalent relationship for the HT parent molecule is
\begin{eqnarray}
2.557 &\leq& E_{\rm rec}^{\rm kin} \leq 3.410 {\rm \ eV,}
\end{eqnarray}
the total internal excitation of the \hhion \ is 0.85~eV, and the translational kinetic energy is 2.557~eV.  We compare these expectations with the calculations of Saenz, Jonsell, and Froelich~\cite{saenz00} in Fig.~\ref{fig:saenzcompare}.

The centroids of the theoretical distributions agree with our expectation but the distributions are not delta functions. Broadening is caused by the fact that atoms in the {\em parent} molecule are always in motion due to thermal and quantum effects, which smears the final-state momentum of the \hethree \  and the momentum of the outgoing leptons.  The calculations of Saenz {\em et al.}~were carried out in the center of mass for \molT \ and HT gas at 30 K; we shall show that, at low temperatures, the chief mechanism for broadening is zero-point motion in the parent molecule. 

\subsection{Spectrum of the electronic ground state}
\label{sec:conceptual_spectrum}

A diatomic molecule at low excitation may be described as a one-dimensional harmonic oscillator:
\begin{eqnarray}
E_v&=&(v+1/2)\hbar \omega_c + a(v\hbar \omega_c)^2; \ v=0, 1, 2, ... \\
\omega_c &=& \sqrt{\frac{k}{\mu}} 
\end{eqnarray}
where $k$ is the force constant for displacements from the equilibrium internuclear separation, and $\mu$ is the  reduced mass.   A small anharmonic term with coefficient $a$ is included.  By fitting the four lowest vibrational states of the H$_2$ molecule \cite{dieke:1958zz} one finds $\hbar \omega_c = 0.5320(5)$ eV and $a=-0.0537(8)$ eV$^{-1}$.  The corresponding value of $\hbar \omega_c$  for \molT \ is then 0.3075 eV,  much larger than $k_BT$ at 30 K (0.003 eV), and also larger than typical rotational excitations (0.005 eV).  In the vibrational ground state, the zero-point motion has an equivalent temperature of about 0.15 eV, or  $\sim 1600$ K, and dominates the line broadening.  The zero-point energy is
\begin{eqnarray}
E_{\rm zp} &\equiv& E_0-E_{-1/2} = \frac{1}{2}\hbar\omega_c - a\left(\frac{1}{2}\hbar\omega_c\right)^2.
\end{eqnarray}

When beta decay occurs, the lepton recoil momentum ${\bf p}$ adds vectorially to the instantaneous momentum ${\bf p_T}$ of the decaying tritium nucleus of mass $m$ within its molecule:
\begin{eqnarray}
{\bf p_f}&=& {\bf p} + {\bf p_T}.
\end{eqnarray}
   The mean kinetic energy of the decaying tritium nucleus is
\begin{eqnarray}
\frac{\left<p_T^2\right>}{2\mu}&=& \frac{1}{2}E_{\rm zp},\\ 
\mu &=& \frac{m_sm}{m_s+m} ,
\end{eqnarray}
and the standard deviation of the excitation energy $E_{\rm exc}$ of the recoil ion is then 
\begin{eqnarray}
\sigma(E_{\rm exc})&=& \frac{p}{m}\sqrt{\frac{1}{3}\left<p_T^2\right> } \\
&=&\sqrt{\frac{p^2}{2m}\left(\frac{2\mu}{3m}E_{\rm zp}\right)} \label{eq:sigma_exc_energy}.
\end{eqnarray}
where $m$ is the mass of the decaying tritium nucleus, and $m_s$ is the nuclear mass of the `spectator' nucleus in the molecule.  For the present purposes we ignore the difference between the nuclear masses of T and $^3$He. 

\begin{figure}[tb]
\includegraphics[width=7.8cm]{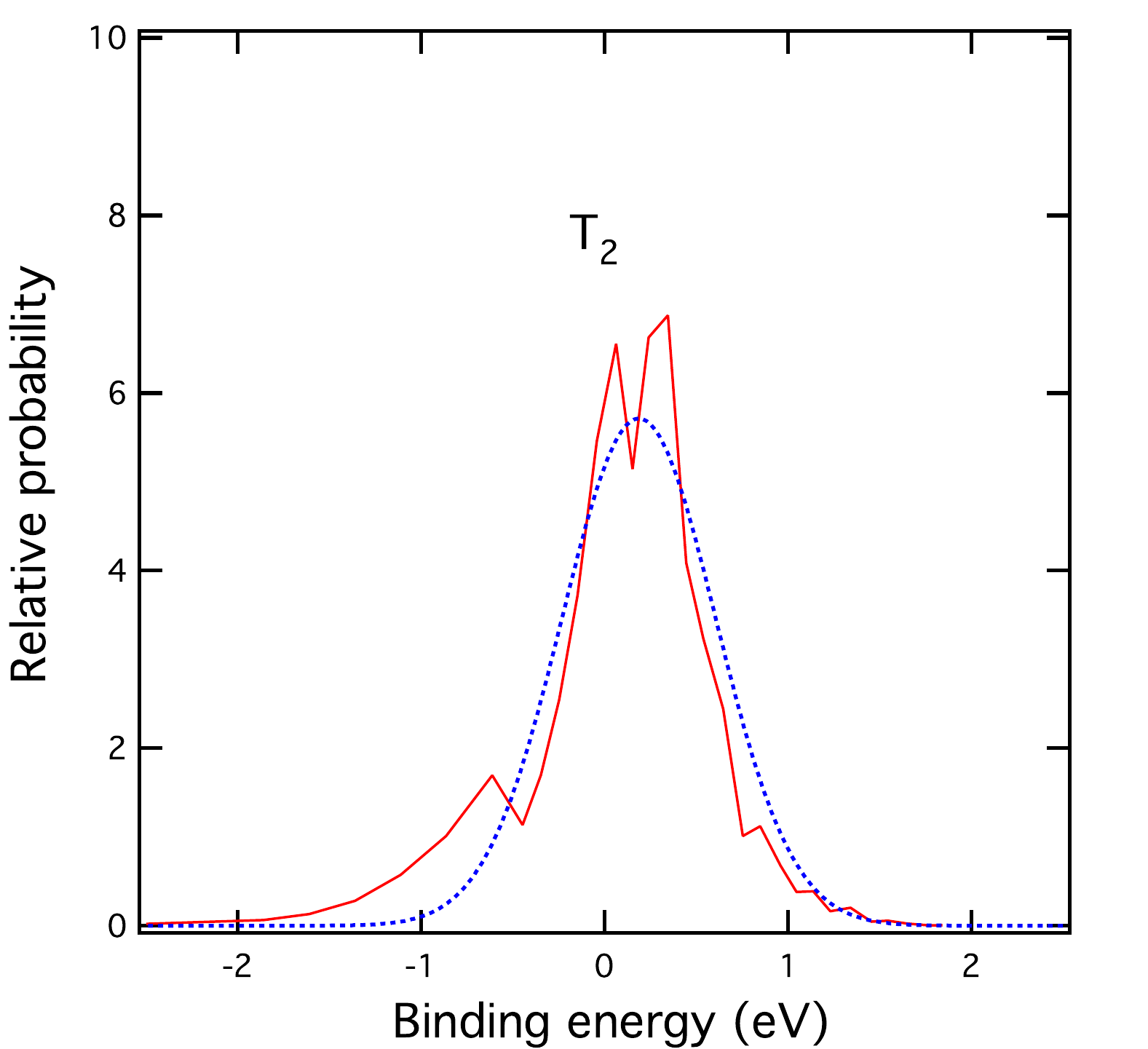}
\includegraphics[width=7.8cm]{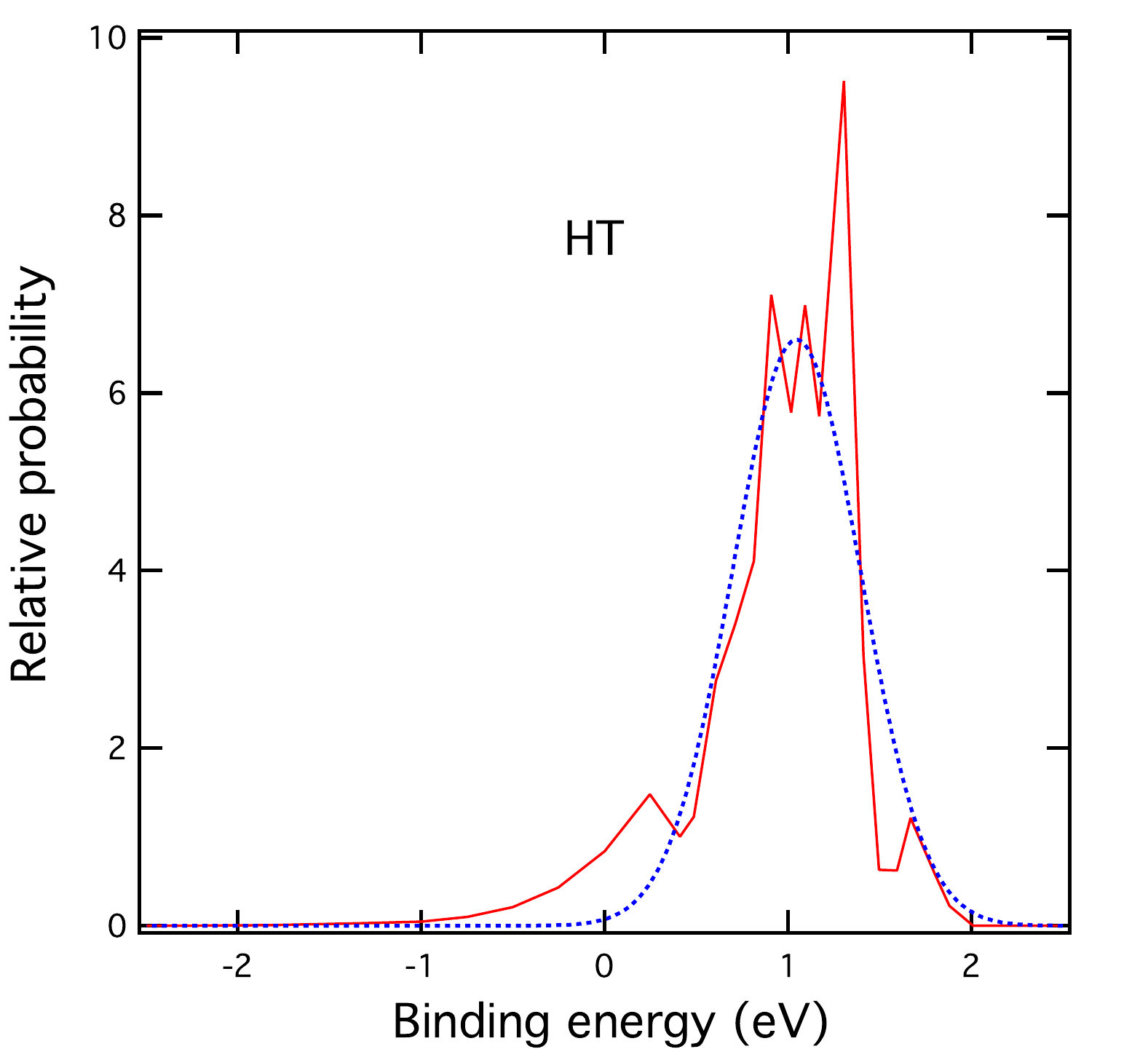}
 \caption{Calculated recoil excitation energy spectra from zero-point motion in the parent molecule (dotted curves, blue online), compared to the final-state distributions  calculated by Saenz {\em et al.}~\cite{saenz00} at 30~K (solid curves, red online). The curves from zero-point motion are parameter-free except for normalization, and have the standard deviations  indicated in Table~\ref{tab:variances}.  An excitation energy of 0 corresponds to a binding energy of 1.897 eV.}
 \label{fig:spectra}
 \end{figure}

Inserting for $E_{\rm zp}$ the relevant zero-point energies for \molT \ and HT, the predicted distributions of recoil excitation energy are compared with the calculated spectra of Saenz {\em et al.}~\cite{saenz00} in Fig.~\ref{fig:spectra}. 
The good agreement (4\%; see  Table~\ref{tab:variances}) underscores the fact that the gross features of the final-state distribution really arise from the {\em initial} state, {\em i.e.} it is mainly  the zero-point  motion of the tritium atom in its molecule that broadens what would otherwise be a line feature.  The broadening  occurs even for the ground-state molecule at absolute zero and is {\em irreducible}.    
Final-state effects assert their presence only through the density of available states in the \molion \ and \hhion \ ions, which modulates the continuous distribution.   That modulation may be calculated by evaluating the overlap integral between the final-state wave functions and the momentum projection operator acting on the initial state as given above.  
 \begin{table}[hbt]
\centering
%Caption above the table
\caption{Root-mean-square widths in eV of the ground-state manifold from the exact calculation of Saenz \etal{}~\cite{saenz00} for initial state $J=0$, and derived from the semiclassical treatment based on the zero-point motion of the parent molecule.}
\label{tab:variances}
\begin{tabular}{lcc}
\hline
Method & \molT\ & \phantom{a} HT  \phantom{a} \\
\hline \hline
Saenz \etal{}~\cite{saenz00} & 0.436  & 0.379  \\
Zero-point motion\phantom{aaa} & 0.420 & 0.354  \\
\hline
\end{tabular}
\end{table}

 Including the smearing effect of zero-point motion, the line feature is broadened to a sufficient extent that a large fraction of the distribution lies above the dissociation threshold, 1.897 eV in \molion.  Jonsell {\em et al.}~\cite{jon99} find that while the intensity of the ground state transitions near the \molT\ beta endpoint amount to 57\% of decays,  18\% absolute lies above the dissociation threshold.  For HT only 1.5\% absolute is above the dissociation threshold  (see Fig.~\ref{fig:spectra}).  Not all such excited states will necessarily dissociate, however, because of the angular momentum barrier for states with high $J$.   Those states may be sufficiently long-lived to decay radiatively.  
  
  The \molT{} vibrational energy interval of 0.308~eV is almost two orders of magnitude larger than the excitation energy $E_{J=1} = 0.00497$~eV of the lowest-lying ortho state (Sec.~\ref{sec:isotopologs}); the zero-point motion is thus the dominant contribution to the final-state width.  If the parent molecule is in an initial state with angular momentum $J$, the root-mean-square width becomes
 \begin{eqnarray}
\sigma(E_{\rm exc})&=&\sqrt{\frac{p^2}{2m}\left(\frac{2\mu}{3m}E_{\rm zp}+\frac{2\alpha^2m_e^2J(J+1)}{3R_0^2m}\right)} \label{eq:sigma_exc_rot},
\end{eqnarray}
where $\alpha$ is the fine structure constant and $R_0$ is the equilibrium internuclear separation in a.u. (1 a.u. = $\hbar/m_e\alpha$).  The variances of the excited-state distributions for \molT{}, DT, and HT for states up to $J=10$ are given in Table~\ref{tab:variancesJ}, and a graphical comparison with the calculations of Doss~\cite{dossphd:2007} up to $J=3$ is displayed in Sec.~\ref{sec:discussion}.

 \begin{table}[hbt]
\centering
%Caption above the table
\caption{Root-mean-square widths in eV of the ground-state manifold of the daughter molecule  from the semiclassical treatment based on the zero-point motion of the parent molecule with the inclusion of rotation.}
\label{tab:variancesJ}
\begin{tabular}{lccc}
\hline
($v,J$) &\phantom{aaaa} \molT\ \phantom{aaaa} & \phantom{aaaa} DT \phantom{aaaa}  & \phantom{aaaa} HT  \phantom{aaaa} \\
\hline \hline
(0,0)&0.4197&0.3972&0.3537\\
(0,1)&0.4331&0.4113&0.3694\\
(0,2)&0.4586&0.4381&0.3991\\
(0,3)&0.4944&0.4755&0.4398\\
(0,4)&0.5385&0.5212&0.4888\\
(0,5)&0.5890&0.5732&0.5439\\
(0,6)&0.6443&0.6299&0.6034\\
(0,7)&0.7035&0.6903&0.6662\\
(0,8)&0.7654&0.7533&0.7313\\
(0,9)&0.8297&0.8185&0.7983\\
(0,10)&0.8956&0.8853&0.8667\\
\hline
\end{tabular}
\end{table}

The objective in a tritium beta-decay experiment is measurement of the neutrino mass via a detailed study of the shape of the electron spectrum near the endpoint.  Energy conservation assures a connection between the molecular final state and the electron energy.  The modification can be directly derived and has a particularly appealing and simple form.

If the tritium atom has a  velocity $\beta_T$ in the center of mass at the instant the decay takes place, the foregoing considerations of zero-point motion in the molecule give 
\begin{eqnarray}
\nonumber \left<\beta_T^2\right> &=&\frac{E_{\rm zp}}{3m}\frac{m_s}{m_s+m} \\
\sigma(E_e) &=& E_e\beta\sqrt{\left<\beta_T^2\right>}.
\end{eqnarray}
 This result is identical to Eq.~\ref{eq:sigma_exc_energy}, the previously derived width for the excitation of the recoil.

\subsection{Recoil energy spectra in dissociation}
\label{sec:concept_recoil_dissoc}

The theory of molecular beta decay can also be used to predict the energy of the ions produced in the decay.  A measurement of the ion energy spectra would be helpful in assessing our understanding of the underlying decay.  As Sec.~\ref{sec:tests_br} discusses in detail, theory predicts that approximately half of the decays of \molT\  and HT lead to dissociative states~\cite{jon99}, whereas experimental data indicate that more than 90\% of the transitions lead to bound molecular ions~\cite{snell57, wex58}.  While there are several plausible experimental and theoretical explanations for this discrepancy, the disagreement motivates an examination of the dissociation-fragment spectrum that would be predicted by  theory.  A future experiment may be able to make a measurement of this spectrum, providing a new test of the theory.

We examine the six dominant electronic configurations in the Born-Oppenheimer approximation as given by Jonsell \etal~\cite{jon99}.  These configurations account for 84\% of the intensity, with the remaining 16\% coming from the electronic continuum.  In the ground-state manifold there is a potential minimum that leads to binding of the \molion\  by almost 2 eV; all electronic excited states are monotonically repulsive with the exception of the first excited state, which has a shallow minimum far outside the Franck-Condon region.   Rotational and vibrational states in the electronic-ground-state manifold are quasibound because of the potential minimum augmented by an angular momentum barrier.  For this analysis we consider these quasibound states to be rotational and vibrational states of a bound (mass-6) ion which dissociates by tunneling through the barrier, analogous to fission.  Conversely, owing to the absence of a binding potential, molecular motion in the electronic excited states corresponds more closely to the unbound scenario in which all the lepton momentum is delivered to a mass-3 recoil ion.   In this case the two fragments  gain additional kinetic energy at dissociation by converting the repulsive potential energy of the excited molecular state at the Franck-Condon spatial separation.   The necessary data for the latter calculation can be found in Fig.~1 of Ref.~\cite{jon99}. 

The laboratory energies of the dissociation fragments from the quasibound ion can be calculated from kinematics.  The laboratory kinetic energy $E_{i \rm (lab)}$ for a fragment of mass $m_i$ is uniformly distributed in the interval
\begin{eqnarray}
E_{i \rm (lab)}&=&\frac{1}{m_i+m_j}\left\{\left(\sqrt{m_iE_{\rm rec}^{\rm kin}}-\sqrt{m_j(E_{\rm exc}-E_B)}\right)^2, \left(\sqrt{m_iE_{\rm rec}^{\rm kin}}+\sqrt{m_j(E_{\rm exc}-E_B)}\right)^2\right\} \nonumber  \\
\label{eq:recoilen}
\end{eqnarray}
for $E_{\rm exc} \ge E_B$ and $m_iE_{\rm rec}^{\rm kin} \ge m_j(E_{\rm exc}-E_B)$, where $m_j$ is the mass of the other fragment and $E_B$ is the binding energy of the molecular ion.  It may be seen from this that the dissociation fragments from the quasibound states do not have translational energies significantly greater than that of the mass-6 ion, $E_{\rm rec}^{\rm kin}$.

Decays populating the electronic excited states produce recoil fragments, at least one of which is itself in an electronic excited state.  Applying the Franck-Condon principle, the electronic excitation energy of the system before dissociation is evaluated at the internuclear separation of the \molT{}, HT, or DT molecule in its ground state, 1.40 a.u.~for all three species \cite{doss06}.  Tables~\ref{tab:excitedstatesT2} and \ref{tab:excitedstatesHT} list the relevant properties for the six dominant electronic states and the kinetic energies of the recoil fragments.
\begin{table*}[tb]
\centering
%Caption above the table
\caption{Structure of excited states and  kinetic energies of dissociation fragments for the decay of \molT.       The probabilities, which are valid in the sudden approximation, are taken from~\cite{doss06} for the case $J_i =  0$, and are very similar for $J_i =  1,2,3$.  The total probability calculated for these six states is 84.2\%.}
\label{tab:excitedstatesT2}
\begin{tabular}{llcccccc}
\hline
State & Asymptotic & Excitation & Excitation  & Total Kinetic &  E($^3$He) & E(T) &  Branch \\
& structure & Energy (eV) & Energy (eV)  & Energy (eV) & eV & eV & (\%) \\
& &  $R=\infty$ & $R=1.40$ a.u. &  & & &  \\
\hline \hline
1 & $^3$He(1s$^2$) + T$^+$ &  $<0$  &  $<1.897$  &  0 &  0 & 0 & 39.0	\\
& $^3$He(1s$^2$) + T$^+$ & $>0$  &  $>1.897$  &  Eq.~\ref{eq:recoilen} &  &  &  18.4	\\
2 & $^3$He$^+$(1s) + T(1s) &	10.981 & 24.587  &  13.606  & $6.8+3.4\eta$ & 6.8 &  17.4	\\
3 & $^3$He(1s2s) + T$^+$ &	20.5	& 31.390 &  10.890 & $5.4+3.4\eta$ & 5.4 &  7.8 \\
4 & $^3$He$^+$(1s) + T(2s+2p) & 21.186  & 36.152 & 14.966 & $7.5+3.4\eta$ & 7.5 &   0.8	\\
5 & $^3$He$^+$(1s) + T(2s-2p) &	21.186  &  36.833 & 15.647 & $7.8+3.4\eta$ & 7.8 &  0.01 \\
6 & $^3$He(1s2p) + T$^+$ & 	21.0	 & 37.513 & 16.513 & $8.3+3.4\eta$& 8.3 &   0.9 \\
\hline
\end{tabular}
\end{table*}

\begin{table*}[tb]
\centering
%Caption above the table
\caption{As Table~\ref{tab:excitedstatesT2}, for the decay of HT. The total probability calculated for these six states is 83.8\%.}
\label{tab:excitedstatesHT}
\begin{tabular}{llcccccc}
\hline
State & Asymptotic & Excitation & Excitation  & Total Kinetic &  E($^3$He) & E(H) &  Branch \\
& structure & Energy (eV) & Energy (eV)  & Energy (eV) & eV & eV & (\%) \\
& &  $R=\infty$ & $R=1.40$ a.u. &  & & &  \\
\hline \hline
1 & $^3$He(1s$^2$) + H$^+$ &  $<0$  &  $<1.897$  &  0 &  0 & 0 & 55.4	\\
& $^3$He(1s$^2$) + H$^+$ & $>0$  &  $>1.897$  &  Eq.~\ref{eq:recoilen} &  &  &  1.5	\\
2 & $^3$He$^+$(1s) + H(1s) &	10.981 & 24.587  &  13.606  & $3.4+3.4\eta$ & 10.2 &  17.4	\\
3 & $^3$He(1s2s) + H$^+$ &	20.5	& 31.390 &  10.890 & $2.7+3.4\eta$ & 8.2 &  7.8 \\
4 & $^3$He$^+$(1s) + H(2s+2p) & 21.186  & 36.152 & 14.966 & $3.7+3.4\eta$ & 11.2 &   0.8	\\
5 & $^3$He$^+$(1s) + H(2s-2p) &	21.186  &  36.833 & 15.647 & $3.9+3.4\eta$ & 11.7 &  0.01 \\
6 & $^3$He(1s2p) + H$^+$ & 	21.0	 & 37.513 & 16.513 & $4.1+3.4\eta$& 12.4 &   0.9 \\
\hline
\end{tabular}
\end{table*}

The {\em total} kinetic energy available to the dissociation fragments by conversion of the interatomic potential in the five excited states is confined to a rather small range between 10 and 17 eV.  An additional amount of kinetic energy $E^\prime_{\rm (trans)} = p^2/2m$ is contributed to the recoil of the beta-decay daughter by the lepton momentum.  We therefore define and use in the Tables a parameter $0\le \eta \le 1$ that is the fraction of the maximum lepton momentum squared.   The other nucleus is a spectator and receives only the kinetic energy obtained from conversion of potential energy.    The maximum energy imparted to a mass-3 daughter recoil is then about 12 eV for  \molT\  and 7 for HT.  The He lines will be broadened by the zero-point  motion as described in Sec.~\ref{sec:conceptual_spectrum}, and all lines will be broadened by the steep gradient of the interatomic potential in the Franck-Condon region.  Moreover, in an experiment the total lepton recoil momentum is not directly measurable; only the electron momentum is.  This introduces a range of values of $\eta$ at each energy, as may be seen in Fig.~\ref{fig:recoilenergy} and Eq.~\ref{eq:recoilenergy}.  A detailed calculation of the line widths is beyond the scope of this work.  

The combination of  the branching ratio to the bound molecular ion and the ion energy spectra provides a complete picture of the decay process.  Measuring the branching ratio and kinematics has the potential to improve our understanding of the efficacy of our current model of molecular beta decay.

\section{Tests of tritium final-state calculations}
\label{sec:tests}

The sub-eV energy scales of the rotational and vibrational excitations and the unknown time scales for further evolution of the final-state products make direct measurement of the final-state distribution, and particularly those aspects that are reflected in the corresponding lepton momentum, all but infeasible.  Of particular concern are detector energy resolution and translational Doppler broadening of the distribution in a real experiment.  The difficulty of a direct measurement has led to a variety of stratagems for indirect verification of the theory. 
In this section we discuss available data from spectroscopy, photodissociation, and mass spectrometry.

\subsection{Studies of the HeH$^+$ molecule}
\label{sec:tests_energy_levels}

\subsubsection{Rotational and vibrational level transitions}
\label{sec:tests_energy_levels_HeH}

Determining the distribution of \molion{} final states populated by beta decay requires calculating the energy levels of \molT{} and of \molion{}. If the same theoretical framework is also applied to calculating the spectra of molecules with other isotopes of He and H, predicted transition energies can be compared against a large number of transition lines measured with high-precision spectroscopic techniques ranging from glow discharge to absorption spectroscopy to Raman spectroscopy. Such a comparison, of course, cannot test the probability of populating each \molion{} state after beta decay, but as we saw in Sec.~\ref{sec:geminal_approach} it has provided valuable input to modern theoretical calculations.

Doss~\cite{dossphd:2007} calculated transition energies between rotational and vibrational levels in the electronic ground state for three tritium-containing parent molecules and for two daughter molecular ions and compared them to published spectroscopic data. For 21 measured transitions in \molT{}~\cite{veirs:1987}, seven in DT~\cite{veirs:1987}, and 12 in HT~\cite{veirs:1987, chuang:1987}, ranging between 120~and 3775~cm$^{-1}$, the theoretical values always agreed within 1~cm$^{-1}$ with a maximum fractional deviation of 0.1\%. For 16 transitions in $^3$HeH$^+$ and 10 in $^3$HeD$^+$~\cite{carrington:1983, crofton:1989, matsushima:1997}, ranging from 71~to 3317~cm$^{-1}$, the agreement is still better, within 0.05\%. However, there do exist experimentally measured transition energies for which no geminal-basis predictions are reported: two rotational-vibrational $Q_1$ transitions in \molT{}~\cite{edwards:1978} and two in DT~\cite{edwards:1979}, three purely rotational transitions in the vibrational ground state of HT~\cite{edwards:1979}, and 12 transitions in hot vibrational bands of HT~\cite{chuang:1987} that fall well outside the energy range of the other measured transitions.

In an earlier calculation in the standard geminal basis, Jonsell~\textit{et al}.~\cite{jon99} predicted transition energies ranging from 598~to 3157~cm$^{-1}$ in helium hydride molecular ions containing the more common isotope $^4$He, allowing validation against a much broader catalog of spectroscopically measured transitions. Five observed transition energies in $^4$HeD$^+$~\cite{carrington:1983} and sixty-two in $^4$HeH$^+$~\cite{tolliver:1979, carrington:1981, bernath:1982, liu:1997, liu:1997a} agree with these predictions to within 0.04\%. The measured widths of seventeen predissociative resonances in $^4$HeH$^+$, $^3$HeH$^+$, $^4$HeD$^+$, and $^3$HeD$^+$~\cite{carrington:1983, liu:1997a} differ from the predicted values by up to an order of magnitude, but the specific machinery for calculating these widths is not used to determine the final-state distribution for neutrino-mass measurements~\cite{jon99}.
No predictions are reported in the geminal basis for 46 additional observed transitions in low-lying vibrational bands of $^4$HeD$^+$~\cite{crofton:1989, matsushima:1997, purder:1992, fan:1998}, or for 36 similar transitions in $^4$HeH$^+$~\cite{crofton:1989, matsushima:1997, purder:1992, liu:1987, blom:1987}.

Despite this great investment of experimental effort, only partial, fragmentary spectra have been measured for these seven molecules. Nonetheless, Coxon and Hajigeorgiu~\cite{coxon:1999} were able to use these data to construct a fitted Born-Oppenheimer potential for the generic molecular helium hydride ion HeH$^+$, and compare it to an {\em ab initio} potential obtained from an older geminal basis with adiabatic corrections from Bishop and Cheung~\cite{bis79}. The two potentials differ by up to 2~cm$^{-1}$ when the nuclei are close together but are in excellent agreement for internuclear distances $R \gtrsim 8$~a.u.; the dissociation energies differ by only 0.27~cm$^{-1}$~\cite{coxon:1999}.  No such comparison has yet been performed for the {\em ab initio} potential based on the most recent geminal basis.

While theoretical predictions for all measured transition energies would be useful, the excellent agreement obtained over 133 transition energies in seven diatomic molecules suggests that the rotational and vibrational energy levels of the electronic ground states are well reproduced in the geminal basis.

\subsubsection{Photodissociation of $^4$HeH$^+$}
\label{sec:photodissoc}

The photodissociation spectrum of $^4$HeH$^+$ may be derived from a sufficiently complete theoretical description of the molecule. Since all electronic excited states of this molecule are dissociative in the Franck-Condon region, one can construct the photodissociation cross section as a function of energy by calculating dipole transitions between the electronic ground state and the electronic excited states. The result depends on the orientation of the internuclear axis relative to the photon polarization vector; the parallel and perpendicular cases must be treated separately. Several other theoretical models ({\it e.g.}~\cite{basu:1984, sodoga:2009}) have been employed to study the photodissociation problem, but have not been applied to neutrino-mass measurements.

The process has been probed experimentally with 38.7-eV (32-nm) photons at the Free-electron LASer in Hamburg (FLASH). The initial measurement~\cite{pedersen:2007} determined the cross section to the $\mathrm{He} + \mathrm{H}^+$ channel, and was not able to define the initial distribution of vibrational states in $^4$HeH$^+$. The second FLASH measurement~\cite{pedersen:2010} incorporated several experimental upgrades to provide additional tests. The $^4$HeH$^+$ beam could optionally be routed through a linear electrostatic ion trap and cooled to the $\nu = 0$ vibrational ground state before being extracted to the interaction region. An improved detection setup, combined with a positive potential across the ion-photon interaction region, allowed the measurement of the branching ratio to the $^4\mathrm{He} + \mathrm{H}^+$ and $^4\mathrm{He}^+ + \mathrm{H}$ channels. In both experiments, the distribution of the initial internuclear axis orientations was assumed to be isotropic. 

Beginning with the same geminal basis set as that used for standard neutrino-mass-relevant calculations, Saenz computed the total photoabsorption cross section assuming that the molecule begins with $\nu = 0$ and is oriented parallel to the photon field~\cite{saenz:2003}.  Dumitriu and Saenz later performed a more detailed calculation in the CI method~\cite{dumitriu:2009} and were able to reproduce those results; despite a 3\% discrepancy in the location of the first resonance, near 25~eV, the two methods are in close agreement at the 38.7-eV energy of the FLASH measurements. CI calculations were also performed for the individual dissociation channels, and for an isotropic molecular orientation, allowing direct comparison with the FLASH cross-section measurement~\cite{pedersen:2007}. The CI calculations give a ratio of $\sim 1.7$ between the two dissociation channels at energies above 35~eV~\cite{dumitriu:2009}, so that the total photoabsorption cross section of $\sim 0.8 \times 10^{-18}$~cm$^2$ at 38.7~eV, predicted in the geminal model~\cite{saenz:2003}, implies a partial cross section of $\sim 0.3 \times 10^{-18}$~cm$^2$ to the $^4\mathrm{He} + \mathrm{H}^+$ channel. The cross-section results, shown in Table~\ref{tab:photodissoc_comp}, demonstrate consistency between experiment and theory, although no theoretical uncertainties have been assigned and the  experimental uncertainty is large.

%Table of the expt vs theory with FLASH
\begin{table}[hbt]
\centering
%Caption above the table
\caption{Photodissociation cross section for $^4$HeH$^+ + \gamma \rightarrow {^4\mathrm{He}} + \mathrm{H}^+$, from geminal and CI theories as well as from an experiment at FLASH. The geminal result, originally computed for both dissociation channels, is corrected for this channel by a factor of 1.7, given by CI calculations.}
\label{tab:photodissoc_comp}

\begin{tabular}{lcc}
\\ \hline
& Molecular & Cross-section  \\
 & Orientation & \phantom{aa}($10^{-18}$\textrm{cm}$^2$)\phantom{aa} \\
\hline \hline
Geminal~\cite{saenz:2003} (with CI~\cite{dumitriu:2009}) & Parallel  & $\sim 0.3$ \\
CI (adiabatic limit)~\cite{dumitriu:2009} & Parallel & $\sim 0.46$ \\
FLASH~\cite{pedersen:2007} & Parallel & $0.4(2)$ \\ 
\hline
CI (adiabatic limit)~\cite{dumitriu:2009} & Isotropic & 1.4 \\
FLASH~\cite{pedersen:2007} & Isotropic & $1.45(7)$ \\ \hline
\end{tabular}
\end{table}

For each event in the FLASH data, the neutral-fragment momentum can be used to reconstruct the initial molecular orientation, under the assumption of fast fragmentation. In general, $\Sigma-\Sigma$ transitions peak for orientations parallel to the field, while $\Sigma-\Pi$ transitions peak when the molecule is oriented perpendicular to the field. For vibrationally cold molecular ions dissociating through the $^4\mathrm{He} + \mathrm{H}^+$ channel, the measured value of $\sim 1:3$ for the $\Sigma:\Pi$ contribution ratio~\cite{pedersen:2010} agrees reasonably well with the CI prediction of $\sim 1:2$~\cite{dumitriu:2009}. There is a clear disagreement in the other channel, however: an experimental measurement of $\Sigma:\Pi \sim 1:1$, compared to a CI prediction of $\sim 1:6$.

Another discrepancy arises in the relative probability of photodissociation to the two channels. For vibrationally cold molecular ions, a ratio of $\sigma_{\mathrm{He}^+ + \mathrm{H}}/\sigma_{\mathrm{He} + \mathrm{H}^+} = 1.70(48)$ was observed in the later FLASH measurement~\cite{pedersen:2010}, in agreement with the prediction of about 1.7 from the CI method~\cite{dumitriu:2009}. However, this ratio was found to drop to $0.96(11)$ when the ion beam was not cooled, contradicting the expectation from the CI potential curves that the ratio would rise. 

Without an error estimation from the theory, the significance of these discrepancies between the CI model and experiment cannot be evaluated.  If the discrepancies hold, they may signal the importance of non-adiabatic effects, which were not included in the calculation of the CI potential curves~\cite{dumitriu:2009}. Such effects are expected to be important to the application of the CI method to the molecular final-state distribution following beta decay in \molT{}.
 
\subsection{Studies of \molion{} and \hhion{} after beta decay}
\label{sec:tests_after_decay}

\subsubsection{Instantaneous final-state distribution after beta decay}
\label{sec:tests_energy_levels_distributions}

In principle, spectroscopy of \molT{} gas can be used to measure the instantaneous population of accessible \molion{} final states after \molT{} beta decay, provided that primary radiative transitions from states excited in beta decay are distinguished from secondary transitions from states excited collisionally. One expects that electronic excitations of \molion{} will dissociate on a time scale of about $10^{-15}$~s, so any observable radiative transitions must arise from  excited dissociation products. Consideration of the dissociation channels for each electronic excited \molion{} state led Jonsell~\textit{et al}.~to conclude that only states representing about 16\% of the total transition probability can result in electronic excited dissociation products that decay via photon emission~\cite{jon99}. A calculation of the full probability distribution of dissociation channels and excitation states is complicated by interference between molecular states and has not been attempted. Experimental data on these transitions are sparse: only one primary transition has been observed in \molT{} spectroscopy, a 468.6-nm line corresponding to the $4s \rightarrow 3p$ transition in \hethree{}$^+$~\cite{wexler:1969, schmieder:1982}.

As seen in Sec.~\ref{sec:tests_energy_levels_HeH}, radiative transitions also occur between rotational and vibrational levels of \molion{}. An infrared emission line ($4.69(3)$~$\upmu$m) has been observed in \molT{} gas and identified as the transition between the $v=1$ and $v=0$ vibrational levels of the \molion{} electronic ground state~\cite{raitzvonfrentz:1974}. The population of excited rotational and vibrational states after \molT{} beta decay depends on the beta momentum, but this experiment did not detect the beta electrons and was therefore insensitive to this variation. The measured excitation probability of the $v=1$ level ($0.4(2)$~\cite{raitzvonfrentz:1974}) thus cannot be compared directly to predictions made near the beta endpoint~\cite{jon99}. 

\subsubsection{Branching ratios to electronic excited final states}
\label{sec:tests_br_elec}

The theory can also be probed by measurements of the branching ratios to various regions of the final-state spectrum following beta decay in \molT{}. A precise measurement of the electron energy spectrum about 25~eV below the endpoint would give the branching ratio to the electronic excited states of \molion{}, which cause a kink in the tritium beta decay spectrum.  With good energy resolution and a large enough sample window, the change in slope can be measured.  The energy resolution must be better than 10~eV to resolve the kink, and the spectrum must be extended to still lower energies to accurately measure the initial slope.  Lower energies correspond to much higher rates, imposing a significant additional burden on the detector system, and corrections for scattering introduce systematic uncertainty.  

Theory predicts that this branching ratio should be about 43\% near the endpoint~\cite{jon99}, but no measurement of the branching ratio to electronic excited states has been reported.   The KATRIN experiment will be able to measure the spectrum in the relevant regime, providing the first direct test of the branching ratio to electronic excited states.  

\subsubsection{First and second moments of FSD from beta decay}

It was pointed out by Staggs {\em et al.}~\cite{Staggs:1989zz} that one of the most direct measures of the accuracy of the FSD is the comparison of the extrapolated endpoint from beta decay with the value expected from mass-spectrometric determinations of the T-$^3$He atomic mass difference, $Q_A$.  If the extrapolated endpoint is obtained from the beta spectrum well below the endpoint, it is the average of the individual quantities $\Delta_{kj}$ and  differs from the ground-state value $\Delta_{00}$ by the first moment of the FSD.  Neglecting neutrino mass and the Heaviside function, which affect the spectrum only at the endpoint, the beta spectrum of Eq.~\ref{eq:betaspectrum} summed over final states $k$ becomes
\begin{eqnarray}
\frac{dN}{dE_e}&\simeq& C F(Z,E_e)\frac{p_eE_e}{\epsilon_0^2}\left(1-\frac{E_e}{M_0}\right)\sum_k\left|W_{k0}\right|^2(\Delta_{k0}-E_e)^2. \label{eq:betaspectrumextrap} 
\end{eqnarray}
The summation may be written in terms of binding energies and the atomic mass difference,
\begin{eqnarray}
& & \sum_k\left|W_{k0}\right|^2\left[(Q_{A}-b_0+2m_e)\left(1-\frac{Q_A-b_0}{2M_0}\right)-m_e+b_{(f)k}-E_e\right]^2 \\
&\equiv& \sum_k\left|W_{k0}\right|^2\left(\delta+b_{(f)k}-E_e\right)^2 
\end{eqnarray}
where terms of order $b_{(f)k}m_e/M_0$ have been dropped and a parameter $\delta$ (the extrapolated endpoint energy for zero final-state binding) has been defined for brevity.  The summation may then be carried out, 
\begin{eqnarray}
\frac{dN}{dE_e}&\simeq& C F(Z,E_e)\frac{p_eE_e}{\epsilon_0^2}\left(1-\frac{E_e}{M_0}\right)\left(\delta+\langle b_{(f)k}\rangle-E_e\right)^2\left(1+\frac{\sigma_b^2}{\left(\delta+\langle b_{(f)k}\rangle-E_e\right)^2}\right) \label{eq:moments}
\end{eqnarray}
The mean binding energy $\langle b_{(f)k}\rangle$ acts as a shift in the extrapolated endpoint $\delta$, and  the variance $\sigma_b^2=\langle b_{(f)k}^2\rangle - \langle b_{(f)k}\rangle^2$ of the (full) binding-energy distribution enters the expression as a shape distortion near the endpoint.  Hence, both the first and second moments of the final-state distribution can be extracted from data for comparison with theory.   Table \ref{tab:moments} lists the first three moments of the binding-energy distributions for two theories. 
\begin{table}[tb]
\centering
%Caption above the table
\caption{Comparison of zeroth, first, and second moments of theoretical final-state distributions \cite{robertson:1988aa}.}
\label{tab:moments}
\medskip
\begin{tabular}{ccccc}
\hline
Reference & \phantom{aa} Energy range  \phantom{aa}  &  \phantom{aa} $\sum_k\left|W_{k0}\right|^2$  \phantom{aa} &  \phantom{aaa} $\langle b_{(f)k}\rangle$  \phantom{aaa} &  \phantom{aaa} $\sigma_b^2$  \phantom{aaa} \\
&    eV  &  &    eV &  eV$^2$ \\
\hline
Fackler {\em et al.} \cite{fackler:1985} & 0 to 165 & 0.9949 & -17.71 & 611.04 \\
Saenz {\em et al.}  \cite{saenz00} & 0 to 240 & 0.9988 & -18.41 & 694.50 \\
\hline
\end{tabular}
\end{table}

In practice, experiments are not analyzed in this way.  Rather, the FSD from theory is used to generate the spectrum to be fitted to data, from which values for $Q_A$ and $m_\nu$ can be extracted.  In addition, only three experiments have used gaseous tritium, and the most modern of these (Troitsk \cite{troitsk:2011}) has a scattering contribution to the spectrum at energies more than 10 eV below the endpoint.  However, the two remaining experiments, LANL \cite{robertson:1991aa} and LLNL \cite{stoeffl:1995aa} used differential spectrometers and magnetic field configurations designed for a broad spectral reach.   The two experiments were in good agreement with each other, but, as is well known, both found an unexpected excess of events in the endpoint region, which is expressed numerically as a negative $m_\nu^2$.  They also yielded concordant values for $Q_A$, but only recently has an accurate determination of $Q_A$ by a non-beta-decay method,  ion cyclotron resonance in the Smiletrap apparatus \cite{0295-5075-74-3-404}, become available  for comparison.  Table \ref{tab:lanlllnl} shows the results of the LANL and LLNL experiments as originally reported, both having been analyzed with the theory of Fackler {\em et al.} \cite{fackler:1985}.  
\begin{table}[tb]
\centering
%Caption above the table
\caption{Atomic mass difference and neutrino mass squared extracted from two experiments, in one case with the original 1985 theoretical calculations of the FSD and in the second case with a more modern calculation.}
\label{tab:lanlllnl}
\medskip
\begin{tabular}{lccl}
\hline
\phantom{aaaa} &  \phantom{aaa} LANL \cite{robertson:1991aa} & LLNL \cite{stoeffl:1995aa}  & \\
\hline
\multicolumn{3}{l}{{\bf As published.} Theory: Fackler {\em et al.} \cite{fackler:1985}} &  \\
 \phantom{aaa} $\Delta_{00}$ & \phantom{aaa} 18570.5(20) & 18568.5(20) &  eV \\
 \phantom{aaa} $Q_A$ &  \phantom{aaa} 18588.6(20) & 18586.6(25) &  eV \\
 \phantom{aaa} $m_\nu^2$ & \phantom{aaa} -147(79) & -130(25) &   eV$^2$ \\
\hline
\multicolumn{3}{l}{{\bf Re-evaluated.} Theory: Saenz {\em et al.} \cite{saenz00}} &  \\
 \phantom{aaa} $\Delta_{00}$ &  \phantom{aaa} 18571.2(20) & 18569.2(20) &  eV \\
 \phantom{aaa} $Q_A$ &  \phantom{aaa}18589.3(20) & 18587.3(25) &  eV \\
 \phantom{aaa} $m_\nu^2$ &  \phantom{aaa} 20(79) & 37(25) &   eV$^2$ \\
\hline
\end{tabular}
\end{table}
The data for those experiments are no longer available, but it is possible to estimate the changes that would be produced with the use of a more modern theory such as that of Saenz {\em et al.} \cite{saenz00} by applying Eqs.~\ref{eq:moments} and \ref{eq:errorest}.  The results are shown in the lower half of the table.  There is excellent agreement between the atomic mass from beta decay and from ion cyclotron resonance, 18589.8(12)  eV, and the large negative value of $m_\nu^2$ is eliminated in both experiments, subject to the limitations of the approximations used.   These results provide a striking measure of experimental confirmation of the calculations of Saenz {\em et al.}, especially in the difficult regime of electronic excited states.

\subsubsection{Branching ratios to molecular and atomic species}
\label{sec:tests_br}

The branching ratio to the bound molecular ion can be extracted from the theory in a straightforward way with certain assumptions.   Two 1950s mass-spectrometry experiments measured this branching ratio for HT~\cite{snell57,wex58}; one of the experiments also measured the branching ratio for \molT{}~\cite{wex58}. The experimental results are consistent with each other but disagree starkly with the theoretical prediction.    

 Calculations of the dissociation likelihood rely on the theoretical dissociation energy of 1.897 eV and assume that all electronic excited states are dissociative, \textit{i.e.}\ there are no fast radiative transitions between the excited states and bound states~\cite{jon99}.  Under these assumptions, and working near the beta endpoint, Jonsell {\em et al.}~\cite{jon99} have calculated a branching ratio to the bound \molion \ molecular ion of $0.39-0.57$, depending on whether the quasibound states above the binding energy dissociate.  An absolute uncertainty of 0.2\%, derived from requiring that the FSD integrate to 100\%, is given for calculation of the entire spectrum but no explicit uncertainties are indicated for the branching ratios. 

A calculation of the differential spectrum as a function of electron energy would permit a more stringent test of the theory than the energy-averaged branching ratio. Experimentally, the ability to distinguish between dissociation products (\textit{e.g}.\ between $^3\textrm{He}^+ + \textrm{T}$ and $^3\textrm{He} + \textrm{T}^+$) allows a stronger test than a simple measurement of the dissociation likelihood, yielding information about how the electronic states are populated after beta decay. 

The first experimental measurement of molecular dissociation following tritium decay was reported for HT by Snell, Pleasanton, and Leming in 1957~\cite{snell57}. The experiment used a mass spectrometer with a conical assembly of ring electrodes that focused ions from an equilibrated mixture of HT, T$_2$, and H$_2$ gas into a magnetic analyzer followed by an electron multiplier~\cite{pleasanton:1957aa}. The measured intensity of the mass-2 peak (H$_2^+$) was used to correct the other peaks for  ionization of the \molT\ or HT gas caused by collisions with beta electrons. The mass-3 peak (T$^+$ or $^3$He$^+$) was corrected for the presence of T$_2$ in the sample gas, based on the ratio of the mass-6 and mass-4 peaks. The correction assumes that HT and T$_2$ have identical dissociation probabilities, which theory does not exclude~\cite{jon99}. The final published result was a $93.2(19)\%$ branching ratio for  HT  decay to the bound $^3$HeH$^+$ ion~\cite{snell57}.

The following year, Wexler used a mass spectrometer with significantly different ion optics to measure the dissociation probability for both  HT and for T$_2$~\cite{wex58}. In this apparatus, the entire source volume was contained within a cone of ring electrodes, which was followed by two distinct deflection stages, one to exclude neutral molecules and one for analysis. A measurement with T$_2$ gas, after correction for an 11.5\% HT impurity, yielded a $94.5(6)\%$ probability of decay to the bound \molion. With a pure sample of HT (0.4\% T$_2$ contamination), the probability of decay to the bound $^3$HeH$^+$ ion was measured at $89.5(11)\%$, in broad agreement ($1.2\sigma$) with the Snell~\textit{et al}.\ measurement~\cite{snell57}. 

In the T$_2$ dataset, the Wexler apparatus was unable to resolve the difference between $^3\textrm{He}^+ + \textrm{T}$ and $^3\textrm{He} + \textrm{T}^+$. For an HT source, however, both Wexler~\cite{wex58} and Snell~\textit{et al}.\ ~\cite{snell57} found that dissociation into a final state of $^3\textrm{He}^+ + \textrm{H}$ was about three times more likely than dissociation into $^3\textrm{He} + \textrm{H}^+$.  This is qualitatively similar to the prediction shown in Table~\ref{tab:excitedstatesHT}, which yields a ratio of 2.1 for the five electronic excited states considered.   

%Table of the expt vs theory
\begin{table}[hbt]
\centering
%Caption above the table
\caption{ Branching ratio to the bound molecular ion for HT and T$_2$.}
\label{tab:dissoc_comp}

\begin{tabular}{cccc}
\hline
\multirow{2}{*}{ Molecule \phantom{aa}} 	& Theory & \phantom{aaa} Snell {\em et al.}\phantom{aaa} &\phantom{aaa} Wexler\phantom{aaa}  \\  \cline{2-4} 
& (Ref.~\cite{jon99}) & (Ref.~\cite{snell57}) & (Ref.~\cite{wex58}) \\
 \hline \hline
HT & 0.55--0.57 & 0.932(19)& 0.895(11) \\ 
T$_2$ & 0.39--0.57 & -- & 0.945(6)\\ \hline
\end{tabular}
\end{table}

Table~\ref{tab:dissoc_comp} summarizes theoretical and experimental results for the branching ratio to the bound molecular ion.  The experimental results for HT and T$_2$ are in stark disagreement with the theoretical predictions.  While a problem of this magnitude with the theory seems unlikely, it is true that geminal calculations of the bound and continuum states are not done in the same basis, and the normalization between the calculations can bias the branching ratio.

To reconcile theory and experiment, other explanations have been advanced for the discrepancy.  The applicability of the theory can be questioned in that the experiments integrated over the entire beta spectrum whereas the sudden approximation is valid when the electron energy is much larger than atomic binding energies.  Another possible mismatch between theory and experiment arises from the evolution of the final state before the ions are detected.  If fast radiative transitions from the electronic excited states to the ground state exist, the experimental measurements would have been too slow to prevent repopulation of the ground state. At the same time, the measurements may have been too fast for some quasi-bound states in the ground-state manifold to dissociate.  The time scales for radiative decays are, however, expected to be orders of magnitude longer than those for dissociation of all but the quasibound states.

A number of experimental issues have also been identified.  The experiments may not have properly accounted for contamination of the mass-6 signal by T$_2^+$ produced via ionization, artificially inflating the measured branching ratio to the bound molecular ion.   This risk was not unknown to the experimenters, who took steps to mitigate it.

Wexler himself favors the explanation that the relative efficiencies between ion species were poorly understood, as the acceptance of both mass spectrometers depended strongly on the initial transverse energy of the ion~\cite{wex58,jon99,ott08}. This transverse energy is dependent on the ion species and can range up to tens of eV following dissociation of excited states of \molion, although most of the dissociation processes should lead to ions in the energy range 3 -- 13 eV.   As computed in Sec.~\ref{sec:concept_recoil_dissoc}, the ion energies resulting from excited-state dissociation are  larger than the $\sim 1$-eV energies for mass-3 fragments in the breakup of the ground state, but whether this accounts for the experimental results is not possible to determine without a model for the acceptance of the mass spectrometers.  A more telling observation, however, is that in the decay of HT the energies of the mass-3 fragments are lower than in the decay of \molT.  That is consistent with Wexler's suggestion because the measured branch to the bound final state HeH$^+$ is smaller than that  to HeT$^+$, perhaps due to better efficiency for detecting the dissociation fragments.   One may also surmise that while dissociation is  energetically allowed  from the ground-state manifold above 1.897 eV excitation, it is strongly hindered by the angular momentum barrier.  A much larger fraction of the HeT$^+$ ground-state manifold can potentially decay this way than for HeH$^+$, and yet the data show the opposite behavior.
 
 The disagreement between theory and experiment has not been satisfactorily explained, although many sources of possible unquantified experimental error have been proposed.  No data are available to test these explanations, however.   Further measurements with the potential to resolve this tension are desirable.

\subsection{Desiderata for a modern experiment}
\label{sec:modern_exp}

A modern dissociation experiment could more closely reproduce the conditions for which the calculations are performed. Detecting the ion in coincidence with a beta electron of measured energy  would allow the experimenter to examine the specific regime where the sudden approximation is valid and to study the variation of the dissociation fraction with electron energy.  The acceptance of the instrument for ions with a range of initial kinetic energies needs to be quantifiable.  Measurement of the ion energy distribution would provide a stronger test of the model.  Complementary information is also available in the coincident photon spectrum but the expected emission falls in the vacuum ultraviolet regime, making it difficult to instrument.  Operating conditions must be such that charge exchange is a minor and quantifiable perturbation. 

A way of implementing many of these objectives is the use of  semiconductor detectors and low-pressure tritium in uniform, coaxial electrostatic and magnetic fields. Mass separation is achieved by time of flight, and the field arrangement offers high efficiency.  When the magnetic field strength is sufficient to collect ions regardless of their transverse momentum, the species-dependent efficiency changes can be eliminated. The radial excursions of the ions can, moreover, be mapped to provide information about their energies and to provide assurance that all have been detected.  Higher detection efficiency allows the source pressure to be lowered, reducing charge exchange, which can artificially lower the measured dissociation probability. An experiment utilizing this approach could more closely reproduce the conditions of the calculations and provide a direct test of specific aspects relevant to the neutrino mass measurement.  Such an experiment, the Tritium Recoil-Ion Mass Spectrometer (TRIMS), is under construction at the University of Washington.

\section{Discussion and Conclusions}
\label{sec:discussion}
\subsection{Impact on tritium neutrino mass experiments}

In this section we aggregate and, where possible, quantify the various ways in which FSD uncertainties contribute when a gaseous tritium source is used to measure neutrino mass.   These fall into 3 groups: theoretical uncertainties in the FSD itself, uncertainties in the degree of temperature equilibration for \molT, and uncertainties in the isotopic composition of the source gas.  

The KATRIN experiment has sufficient statistical power that data-taking can be concentrated in the last 20 eV of the spectrum, which removes the theoretical uncertainties in electronic excitation of the molecule as a major concern.  There is remaining uncertainty in the width of the ground-state manifold of rotational and vibrational excitations, but we have shown that  the broadening has a very simple origin, mainly zero-point motion.  Indeed, the semiclassically derived analytic expression yields a variance that agrees with the full theoretical calculation to 7\%.  Beyond this, a quantitative uncertainty estimate is lacking, and knowledge of  the variance at  the 1\% level has been assumed in the design of experiments like KATRIN.  We have reviewed a variety of tests of the theory, finding generally excellent agreement, with the one serious exception being the branching ratio to the bound mass-6 ground state manifold.   A new experiment would provide substance for a re-evaluation of the theoretical uncertainties.

An accurate characterization of the composition of the source is necessary for KATRIN.  The source gas is high-purity \molT.  To determine the isotopic composition, the KATRIN collaboration has developed a laser Raman spectroscopy system called LARA.  This system has achieved 0.1\% precision~\cite{fischer:2011aa} and better than 10\% accuracy~\cite{schlosser:2013} in measurements of the isotopic composition. In principle,  a laser Raman system can also provide information about the ortho-para ratio.  However, due to the difficulty of {\em in situ} measurement, LARA is located at a high-pressure stage prior to cooling and injection into the source.  The KATRIN collaboration is studying an extension of the LARA system to measure the ortho-para ratio and is conducting ongoing simulation work on the evolution of the ortho-para ratio and other source parameters.

The KATRIN windowless, gaseous tritium source vessel will be maintained at a temperature of 30~K.    In thermal equilibrium at this temperature the ortho-para ratio is approximately 1:1 and states with $J > 1$ are not appreciably populated. The time each molecule spends in the cooled source, however, is short compared to the spin relaxation time.  The ortho-para ratio of the gas within the source is therefore expected to be close to 3:1.

Disequilibrium in the source is not confined to the ortho-para ratio because depopulation of higher excited states in free space requires quadrupole transitions that are very slow.  The de-excitation process is therefore predominantly collisional and apparatus-dependent.  Incomplete thermalization of these excited states would be a source of uncertainty if undiagnosed.

These sources of uncertainty in the FSD translate directly to an uncertainty in the neutrino mass-squared.   Robertson and Knapp~\cite{robertson:1988aa} have shown that any neglected contribution to the variance of the FSD, $\Delta \sigma_{\rm FSD}^2$, modifies the extracted neutrino mass-squared by
\begin{eqnarray}
\Delta m_\nu^2 & \simeq & -2 \Delta \sigma_{\rm FSD}^2. \label{eq:errorest} 
\end{eqnarray}

Doss \etal~\cite{doss06} calculated the final state distributions arising from the lowest four rotational states of \molT{} and the lowest two states of HT and DT, {\em i.e.}\ those populated in a 30-K thermal source.   The FSDs were binned with 0.01-eV resolution compared to the 0.1-eV resolution used in reporting previous results~\cite{saenz00}.  We have estimated the variance of each binned distribution in two ways: using the central bin energy value and the reported mean energy value.  We take the average of the two results as the best estimate of the variance and half the difference as the width (standard deviation) of the error distribution.  The mean excitation energies and estimated variances of the FSDs are listed in Table~\ref{tab:excvar}.  Unfortunately the distributions for higher rotational states of \molT{} were not available, and distributions for HT are not available with the required binning resolution.  Future calculations of the FSD, such as calculations using the configuration-interaction method, will be helpful in expanding and improving the estimates of the variances. 

\begin{figure}[tb]
\includegraphics[width=10cm]{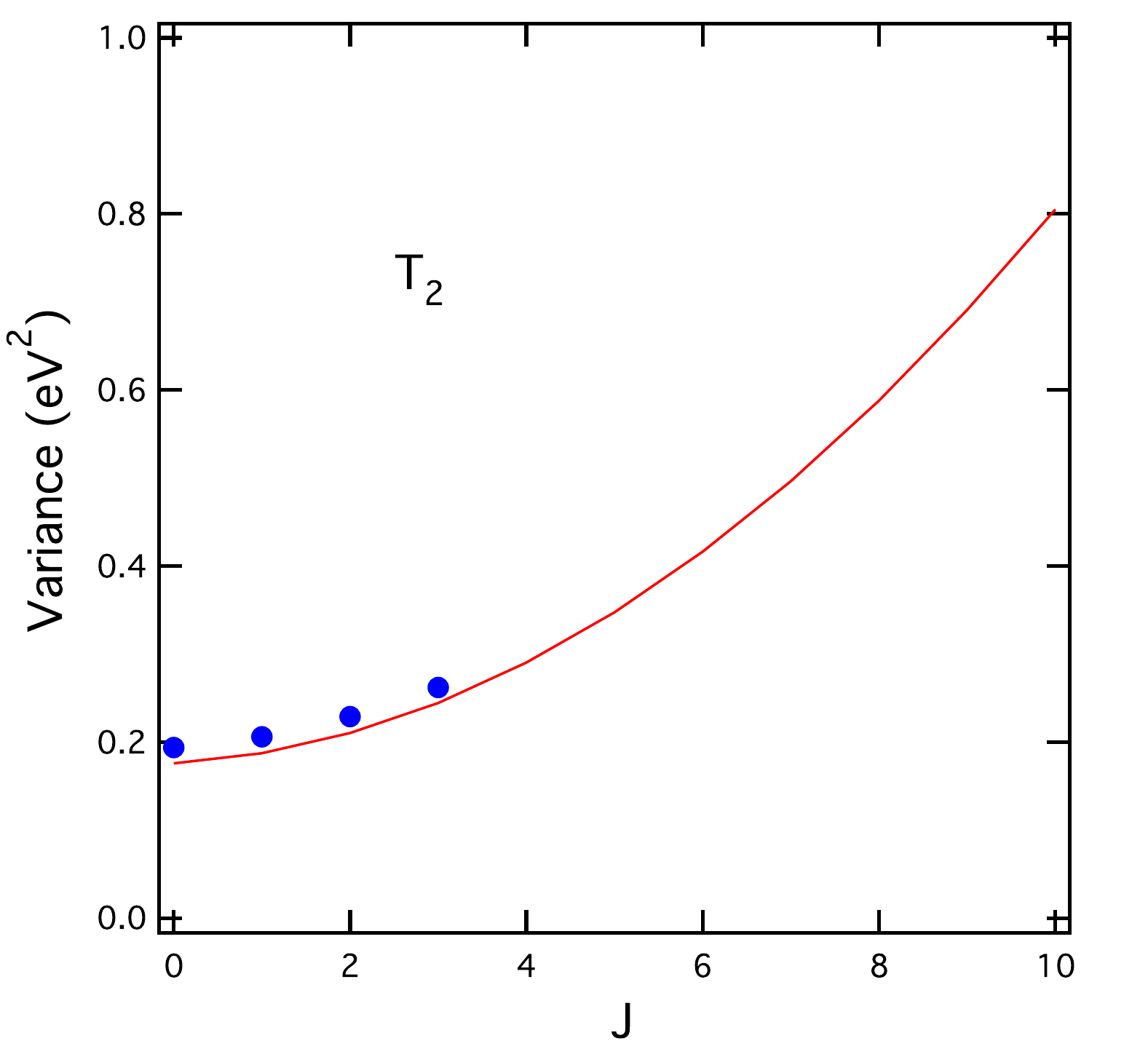}
 \caption{Comparison of the variance of the ground-state-manifold FSD produced in \molT\ decay as calculated in the semiclassical model, Eq.~\ref{eq:sigma_exc_rot} (solid curve, red online), with variances taken from calculations for states up to $J=3$ described in Ref.~\cite{doss06}  (blue dots).}
 \label{fig:variances}
 \end{figure}

Figure~\ref{fig:variances} compares the semiclassical variances calculated for initial states ($0,J$) in \molT\ using Eq.~\ref{eq:sigma_exc_rot} to the variances estimated from the calculations of Doss \etal\ \cite{doss06}.  From the figure we conclude that the semiclassical model is a good proxy for the FSD variance.  The difference between the two is about 7\% and independent of J.  Of this difference, 1\% is attributable to our more accurate result for $E_{{\rm rec, max}}^{\rm kin}$ because  all contributions to the variance are proportional to $p^2/2m$.   Given the limited set of full FSD calculations available, we use the semiclassical variances to estimate the systematic errors associated with various experimental parameters. 
\begin{table}
\centering
\caption{Mean excitation energy and variances extracted from the FSD calculations of reference~\cite{dossphd:2007}.  There is a small contribution to the variance ($<0.004$ eV$^2$) from binning. \label{tab:excvar}}
\medskip
\begin{tabular}{cccc}
\hline
Source			&	\phantom{aaa} $J$ \phantom{aaa} & \phantom{aa}Mean $E_\textrm{exc}$ (eV)\phantom{aa}	& 	$\sigma_{J}^2$ (eV$^2$)  \\ 
\hline 
\hline
\multirow{4}{*}{\molT{}}	&	0	& 	1.752	&	0.194 \\
					&	1	&	1.751	&	0.206 \\
					&	2	&	1.750	&	0.215 \\
					&	3	&	1.749	&	0.262 \\ 
\hline
\multirow{2}{*}{DT}		&	0	&	1.752*	&	0.175 \\
					&	1	&	1.752*	&	0.188 \\ 
\hline
\multicolumn{4}{l}{\footnotesize *Shifted to compensate for different recoil kinetic energy~\cite{dossphd:2007}.}
\end{tabular}
\end{table}

 After shifting the excitation energy to compensate for differences in the recoil kinetic energy, the effective mean excitation energy of each of the FSDs corresponds to the same laboratory endpoint energy for each isotopolog.  Thus the variance of the summed distribution can be taken as the sum of the variances for each isotopolog $i$ and each rotational state $J$, weighted according to their populations $f_i$ and $P_{J,i}$ for isotope and rotational state, respectively.   An additional variance contribution arises from the translational Doppler broadening $\sigma^2_{\rm trans}$ at a given temperature $T$.    The overall variance $\sigma_{\rm tot}^2$ of the line broadening can be derived:
\begin{eqnarray}
\sigma_{J,i}^2 &=& \frac{p^2}{2m}\left[\frac{2\mu_i}{3m}E_{{\rm zp}(i)}+\frac{2\alpha^2m_e^2J(J+1)}{3R_0^2m}\right] \label{eq:varj}\\
\sigma_{\textrm{FSD},i}^2 &=& \sum_{J} P_{J,i}\sigma_{J,i}^2  \label{eq:varfsd}\\
\sigma^2_{\textrm{trans},i} &=& \frac{p^2}{2m}\frac{2mk_BT}{m_{s,i}+m} \label{eq:vartrans}\\
\sigma_{\rm tot}^2 &=& \sum_{i} f_{i}\left( \sigma_{\textrm{FSD},i}^2 + \sigma^2_{\textrm{trans},i}\right) \label{eq:var} 
\end{eqnarray}
The $P_{J,i}$ weights are given by a Boltzmann distribution for the  temperature $T$.  (The translational and rotational temperatures need not be the same).  The probability distribution is calculated independently for each isotopolog and summed according to the activity fraction $f_i$ of each isotopolog in the source.    The source activity may be expressed in terms of a parameter $\epsilon_\textrm{T}$ that is the equivalent fractional activity of the gas compared to pure \molT.    Additionally the ratio of HT to DT in the source gas $ \kappa = f_{\textrm{HT}}/f_{\textrm{DT}}$ is used to characterize the makeup of the active contaminants.  Eq.~\ref{eq:isotopicweight} shows the functional form of the isotopic weights.  
\begin{equation} \label{eq:isotopicweight}
   f_i = \left\{
     \begin{array}{ll}
       2\epsilon_\textrm{T}-1  &, i= \textrm{T}_2\\
       2(1-\epsilon_\textrm{T})/(1+\kappa) &, i = \textrm{DT}\\
       2(1-\epsilon_\textrm{T})\kappa/(1+\kappa) &, i = \textrm{HT}	
     \end{array}
   \right.
\end{equation}
Neglecting inert isotopologs H$_2$, HD, and D$_2$,  $\epsilon_\textrm{T}$  is confined to the range $0.5\le\epsilon_\textrm{T} \le 1$ and  is assigned a reference value of 0.95 as in the KATRIN Design Report~\cite{katrin04}.    The reference value of $\kappa$ is taken to be 0.1 because the fractional distillation process results in higher levels of deuterium than of protium.

 \begin{table}[hbt]
\centering
%Caption above the table
\caption{Rotational-state distributions for \molT\ at 30~K and 300~K.  The energies are those used in Ref.~\cite{dossphd:2007} and variances are from the semiclassical width, Eq.~\ref{eq:sigma_exc_rot}.  Probabilities $P$ are calculated from the partition function (Eq.~\ref{eqn:partition}) using the energies listed in the table and the contributions to the total FSD variance are computed accordingly.}
\label{tab:jstatesT2}
\begin{tabular}{crcrcrcrc}
\hline
\phantom{a}$J$ \phantom{a} & \phantom{a} $ E_J$  &  \phantom{a} $\sigma^2_{J, {\rm T}_2}$ \phantom{a}  &\multicolumn {2}{c} {30 K, Thermal} & \multicolumn {2}{c} {300 K, Thermal} &\multicolumn {2}{c} {30 K, $\lambda = 0.75$}\\ \cline{4-9} 
  &	(meV)			& (eV$^2$)	&\phantom{aa}P (\%) 	& Var Contr  & \phantom{aa}P (\%) 	& Var Contr  &P (\%) 		& Var Contr\\

\hline \hline
0	&	0.00	&	0.1762	&	43.70	&	0.0768	&	4.73	&	0.0083	&	24.6	&	0.0434	\\
1	&	5.01	&	0.1875	&	55.70	&	0.1040	&	35.00	&	0.0656	&	75.0	&	0.1410	\\
2	&	15.02	&	0.2103	&	0.62	&	0.0013	&	13.20	&	0.0277	&	0.35	&	0.0007	\\
3	&	30.05	&	0.2445	&	0.01	&	0.0000	&	30.70	&	0.0752	&	0.01	&	0.0000	\\
4	&	50.08	&	0.2900	&	0.00	&	0.0000	&	6.03	&	0.0175	&	0.00	&	0.0000	\\
5	&	75.11	&	0.3469	&	0.00	&	0.0000	&	8.33	&	0.0289	&	0.00	&	0.0000	\\
6	&	105.16	&	0.4152	&	0.00	&	0.0000	&	1.02	&	0.0042	&	0.00	&	0.0000	\\
7	&	140.21	&	0.4949	&	0.00	&	0.0000	&	0.90	&	0.0045	&	0.00	&	0.0000	\\
8	&	180.27	&	0.5859	&	0.00	&	0.0000	&	0.07	&	0.0004	&	0.00	&	0.0000	\\
9	&	225.34	&	0.6883	&	0.00	&	0.0000	&	0.04	&	0.0003	&	0.00	&	0.0000	\\
10	&	275.42	&	0.8022	&	0.00	&	0.0000	&	0.00	&	0.0000	&	0.00	&	0.0000	\\ \cline{5-5} \cline{7-7} \cline{9-9}
	&		&	\multicolumn{2}{r}{FSD Variance}			&	0.1830	&		&	0.2330	&		&	0.1850	\\
\hline
\end{tabular}
\end{table}

 \begin{table}[hbt]
\centering
%Caption above the table
\caption{Rotational-state distributions for DT at 30~K and 300~K.  The energies and variances are from the semiclassical model (see Eq.~\ref{eq:sigma_exc_rot}).  Probabilities are calculated from the partition function (Eq.~\ref{eqn:partition}) using the energies listed in the table and the contributions to the total FSD variance are computed accordingly.}
\label{tab:jstatesDT}
\begin{tabular}{crcrcrc}
\hline
\phantom{a}$J$ \phantom{a} & \phantom{a} $ E_J$  &  \phantom{a} $\sigma^2_{J, {\rm DT}}$ \phantom{a}  &\multicolumn {2}{c} {30 K, Thermal} & \multicolumn {2}{c} {300 K, Thermal} \\ \cline{4-7} 
  &	(meV)			& (eV$^2$)	&P (\%) 		& Var Contr \phantom{aa} &P (\%) 		& Var Contr\\
%\phantom{a}$J$ \phantom{a} & $\phantom{a} E_J$ (meV)&  $\sigma^2_{J, {\rm DT}}$ (eV$^2$) &\multicolumn {2}{c} {30 K Thermal} & \multicolumn {2}{c} {300 K Thermal} \\ \cline{4-7} 
%  &				& 	&P (\%) 		& Var Contr &P (\%) & Var Contr 	\\
\hline \hline
0	&	0.00	&	0.1578	&	78.70	&	0.1242	&	11.61	&	0.0183	\\
1	&	6.25	&	0.1692	&	21.02	&	0.0356	&	27.36	&	0.0463	\\
2	&	18.76	&	0.1919	&	0.28	&	0.0005	&	28.11	&	0.0540	\\
3	&	37.52	&	0.2261	&	0.00	&	0.0000	&	19.05	&	0.0431	\\
4	&	62.53	&	0.2716	&	0.00	&	0.0000	&	9.31	&	0.0253	\\
5	&	93.80	&	0.3285	&	0.00	&	0.0000	&	3.39	&	0.0111	\\
6	&	131.32	&	0.3968	&	0.00	&	0.0000	&	0.94	&	0.0037	\\
7	&	175.09	&	0.4765	&	0.00	&	0.0000	&	0.20	&	0.0010	\\
8	&	225.12	&	0.5675	&	0.00	&	0.0000	&	0.03	&	0.0002	\\
9	&	281.40	&	0.6700	&	0.00	&	0.0000	&	0.00	&	0.0000	\\
10	&	343.93	&	0.7838	&	0.00	&	0.0000	&	0.00	&	0.0000	\\ \cline{5-5} \cline{7-7}
	&		&	\multicolumn{2}{r}{FSD Variance}			&	0.1603	&		&	0.2029	\\
\hline
\end{tabular}
\end{table}

 \begin{table}[hbt]
\centering
%Caption above the table
\caption{Rotational-state distributions for HT at 30~K and 300~K.  The energies and variances are from the semiclassical model (see Eq.~\ref{eq:sigma_exc_rot}).  Probabilities are calculated from the partition function (Eq.~\ref{eqn:partition}) using the energies listed in the table and the contributions to the total FSD variance are computed accordingly.  }
\label{tab:jstatesHT}
\begin{tabular}{crcrcrc}
\hline
\phantom{a}$J$ \phantom{a} & \phantom{a} $ E_J$  &  \phantom{a} $\sigma^2_{J, {\rm HT}}$ \phantom{a}  &\multicolumn {2}{c} {30 K, Thermal} & \multicolumn {2}{c} {300 K, Thermal} \\ \cline{4-7} 
  &	(meV)			& (eV$^2$)	&P (\%) 		& Var Contr \phantom{aa} &P (\%) 		& Var Contr\\
\hline \hline
HT													
0	&	0.00	      	&    0.1251	&    94.09	&	0.1177	&	18.12	&	0.0227	\\
1	&	10.00	&	0.1365	&	5.91	&	0.0081	&	36.93	&	0.0504	\\
2	&	29.99	&	0.1592	&	0.00	&	0.0000	&	28.40	&	0.0452	\\
3	&	59.98	&	0.1934	&	0.00	&	0.0000	&	12.46	&	0.0241	\\
4	&	99.97	&	0.2389	&	0.00	&	0.0000	&	3.41	&	0.0082	\\
5	&	149.95	&	0.2958	&	0.00	&	0.0000	&	0.60	&	0.0018	\\
6	&	209.94	&	0.3641	&	0.00	&	0.0000	&	0.07	&	0.0003	\\
7	&	279.91	&	0.4438	&	0.00	&	0.0000	&	0.01	&	0.0000	\\
8	&	359.89	&	0.5348	&	0.00	&	0.0000	&	0.00	&	0.0000	\\
9	&	449.86	&	0.6373	&	0.00	&	0.0000	&	0.00	&	0.0000	\\
10	&	549.83	&	0.7511	&	0.00	&	0.0000	&	0.00	&	0.0000	\\ \cline{5-5} \cline{7-7}
	&		&	\multicolumn{2}{r}{FSD Variance}			&	0.1258	&		&	0.1526	\\
\hline
\end{tabular}
\end{table}

Table~\ref{tab:jstatesT2} shows the rotational-state distributions for \molT\ thermal 30~K, thermal 300~K, and nonthermal 30~K ($\lambda = 0.75$) sources along with the semiclassical FSD variances.  Also shown is the contribution each state makes to the total FSD variance of the source in each configuration.  The rotational-state distributions for DT and HT are shown in Tables~\ref{tab:jstatesDT} and \ref{tab:jstatesHT}, respectively.  (The rotational-state energies differ slightly from those given by Doss~\cite{doss06}, possibly because centrifugal stretching is not included here.) The rotational states up to $J=7$ contribute significantly at room temperature and further work is necessary to provide an accurate assessment of the systematic error associated with the experimental uncertainty in the rotational-state distribution.  Measurement of the rotational-state temperature and calculations of the higher rotational-state FSDs would significantly improve the error estimates.

To quantify the impact of using an incorrect FSD to analyze neutrino-mass data we examine the differences in variances that arise due to changes in temperature, isotopic purity and ortho-para conditions.  For small deviations from the operating parameters the corresponding error in the extracted neutrino mass-squared can be derived from Eq.~\ref{eq:errorest}.  Below, we derive the functional form for ortho-para ratio errors, temperature fluctuations and errors in the isotopic composition.  The nominal source parameters are shown in Table~\ref{tab:KATRIN-parameters}.

\begin{table}
\centering
\caption{Reference values of parameters used in estimating FSD and Doppler contributions to the projected uncertainty in the extracted $m_\nu^2$ for KATRIN.}
\label{tab:KATRIN-parameters}
\begin{tabular}{ll}
\hline 
Parameter \phantom{aaaaaaaaaaaaaaaa}		& Value		\\
\hline \hline
Source temperature 					& $T = 30$~K 				\\
Ortho fraction 							& $\lambda = 0.75$		\\
Tritium fraction in WGTS					& $\epsilon_\textrm{T} = 0.95$		\\
Ratio of DT to HT						& $\kappa = 0.1$ 				\\
\hline
\end{tabular}
\end{table}

The temperature of the source is a key parameter determining the width of the final state distribution.  As previously stated, the rotational states of homonuclear \molT\ do not equilibrate on short time scales~\cite{souers} and the exact time required for thermalization in the KATRIN source depends not only on the gas density but also on the materials the gas contacts ({\em i.e.} walls, permeators, etc.).  The temperature changes the initial rotational-state distribution of the source as seen in the partition function.  For small fractional changes in temperature  the exponential factors can be expanded, and the resulting shift in variance can be expressed in terms of the fractional temperature change.  For a cryogenic source only the $J=0$ and $J=1$ states contribute significantly and the shift in FSD variance for a given isotopolog simplifies to a single term, which may be written:
\begin{eqnarray}
\delta \sigma^2_{\textrm{FSD},i} &=& \sum_{J} \sigma_{J,i}^2 P_{J,i}  \sum_{n} P_{n,i}  \frac{E_{J,i}-E_{n,i}}{kT}   \phantom{a}\frac{\delta T}{T}\\
&\approx& P_0 P_1  \frac{E_1}{kT^2} (\sigma_1^2 - \sigma_0^2) \delta T.
\end{eqnarray}
Table~\ref{tab:doppler} shows the  translational Doppler variance temperature-variation coefficients for \molT{}, DT and HT, computed from:  
\begin{equation}\label{eq:doppler}
\delta \sigma^2_{\textrm{trans},i} =  \frac{p^2}{2m}\frac{2mk_B}{m_{s,i}+m} \delta T.
\end{equation}

\begin{table}
\caption{Variation with temperature of the translational Doppler contribution to the variance for a source near 30~K, calculated from Eq.~\ref{eq:doppler}.}
\label{tab:doppler}
\medskip
\begin{tabular}{cc}
\hline
Source \ & $\frac{\delta \sigma^2_\textrm{trans}}{\delta T} \phantom{a} (10^{-3}~\textrm{eV}^2/\textrm{K})$  \\ \hline \hline
\molT{} & 0.147 \\
DT & 0.176\\
HT& 0.220 \\
\hline
\end{tabular}
\end{table}

The shifts in variance due to the FSD and translational effects are additive, and each contributes to the overall shift in the extracted neutrino mass-squared according to Eq.~\ref{eq:errorest}. A temperature change of 0.15~K from the nominal 30~K results in a shift in extracted neutrino mass-squared of $0.11\times10^{-3}\textrm{ eV}^2$.   

In reality both thermal fluctuations and inaccuracy in the measurement of the temperature contribute to the uncertainty on neutrino mass.  It is reasonable to assume these are uncorrelated errors and thus two independent thermal factors appear in the error budget.  The expected temperature fluctuations and uncertainties are taken from the work of Grohmann {\em et al.}~\cite{Grohmann2011438,grohmann:2013aa}.

The isotopic purity of the source plays a major role in neutrino-mass experiments because the width of the FSD varies significantly between isotopologs.   In addition to the dependence on the  tritium activity fraction $\epsilon_\textrm{T}$, there is a dependence on the relative population $\kappa$ of contaminants HT and DT.  Tables~\ref{tab:jstatesT2},~\ref{tab:jstatesDT}~and~\ref{tab:jstatesHT} show the variance of the distribution for 30-K sources of tritium-containing isotopologs.  The \molT\ results include the thermal source as well as the nonthermal source with $\lambda = 0.75$.
The large differences in the FSD variances between HT, DT and \molT{} demonstrate the importance of knowing the isotopic purity.   The shift in the variance that occurs when the tritium purity of the source $\epsilon_\textrm{T}$ changes can be written 
\begin{equation}\label{eqn:isotopicpurity}
\delta \sigma^2 = \bigg[ 2\sigma_{{\rm T}_2}^2 - \frac{2}{1+\kappa}\sigma_{\rm DT}^2 - \frac{2\kappa}{1+\kappa}\sigma_{\rm HT}^2 \bigg]\delta\epsilon_\textrm{T}, 
\end{equation}
where $\sigma^2_i$ is the sum of the FSD (Eq.~\ref{eq:varfsd}) and translational (\ref{eq:vartrans}) terms.  Similarly, the dependence on $\kappa$ takes the form

\begin{equation}
\delta \sigma^2 = \frac{2(1-\epsilon_\textrm{T})}{(1+\kappa)^2}\bigg[-\sigma_{\rm DT}^2+\sigma_{\rm HT}^2\bigg]\delta \kappa.
\end{equation}

Starting from the nominal source parameters (Table~\ref{tab:KATRIN-parameters}) and introducing an uncertainty of 1\% on the atomic purity of the source would lead to a uncertainty on the neutrino mass-squared of $0.96\times10^{-3}\textrm{ eV}^2$.   While conflicting previous results have led to confusion over the impact of errors in the measurement of isotopic purity~\cite{doss06,schlosser:2013}, our results are consistent with the earlier published work of Doss \etal~\cite{doss06} which concluded that it plays a major role.  

The impact of the ortho-para condition of the source can also be derived from Eq.~\ref{eq:var} by considering a slight reordering of rotational states.  Due to the two-state nature of the homonuclear system, the state distribution for \molT\ is often separated out in terms of the even (para) and odd (ortho) states.  The sum of probabilites for all the odd states is the ortho fraction of the source: 
\begin{equation}
\lambda = \sum_{J \textrm{ odd}} P_J.
\end{equation}

The variances of the ortho and para states can then be considered separately and even normalized independently to yield ortho and para state probabilities, labeled $P_{\textrm{ortho},J}$ and $P_{\textrm{para},J}$ respectively.  The total variance is then the sum of two states weighted according to the $\lambda$ factor.  
\begin{eqnarray}
\sigma_{\textrm{FSD,\molT{}}}^2 &=& \lambda \sum_{J \textrm{ odd}} P_{\textrm{ortho},J} \sigma_{\textrm{J}}^2 +(1-\lambda)\sum_{J \textrm{ even}} P_{\textrm{para},J}\sigma_{\textrm{J}}^2\\
&\equiv&\lambda \sigma^2_\textrm{ortho} + (1-\lambda)\sigma^2_\textrm{para}.
\end{eqnarray} 

If the probabilities within the ortho (para) state relative to the other states are not changing then the impact of the ortho-para transitions can be assessed in terms of the independent ortho and para state variances.  Under this assumption, the dependence on $\delta \lambda$ is simply characterized by the difference in the FSD variances arising from the ortho and para distributions:  
\begin{equation}
\delta \sigma_{\textrm{FSD}}^2 = (\sigma_{\textrm{ortho}}^2 - \sigma_{\textrm{para}}^2) \delta \lambda.
\end{equation}

For cryogenic sources the equation of the shift in neutrino mass-squared further simplifies, only depending on the difference in the variances of the $J=0$ and $J=1$ states.  For small changes in temperature which do not appreciably change the occupation of the higher states, the shift in variance is independent of temperature.  The contributions from DT and HT remain unchanged as ortho-para considerations only apply to the homonuclear isotopolog.  The effect of a change in ortho-para ratio on the extracted neutrino mass-squared is given by:
\begin{equation}
|\Delta m_\nu^2| \sim 2(2\epsilon_\textrm{T} -1)(\sigma_{J=1}^2-\sigma_{J=0}^2)\delta \lambda.
\end{equation}

Given the relatively short time that molecules will spend at cryogenic temperatures in the KATRIN source, the ortho fraction is expected to be close to 0.75, corresponding to the 700~K permeator through which the gas passes in atomic form.   A lower bound of 0.57 is set by the beam-tube temperature of 30~K.  If $\lambda$ lies at an unknown value between these bounds the corresponding uncertainty on the extracted neutrino mass-squared would be $3.8\times 10^{-3}\textrm{ eV}^2$.  Fortunately this is not expected to be the case and early simulations  indicate that even in pessimistic scenarios only 3\% of the ortho source molecules will transition from the ortho state to the para state~\cite{kleesiek:2014}.  These KATRIN simulations show  a shift in neutrino mass-squared of $0.48(7) \times 10^{-3}\textrm{ eV}^2$ due to ortho-para transitions.  Our calculation is $0.44\times10^{-3}$~eV$^2$, in good agreement with the results of the simulation.  Thus under standard scenarios the ortho-para ratio is not expected to contribute significantly to the uncertainty on the neutrino mass-squared.  

While not considered a significant concern for KATRIN, from an experimental perspective the ortho-para ratio warrants more study as the $\lambda$ factor and associated systematic error can potentially be measured.  Current work by the LARA subgroup of KATRIN focuses on how to measure the ortho-para ratio using a modified version of the setup used to measure the isotopic ratio.  

Table~\ref{tab:KATRIN-FSDerrors} summarizes the projected role of molecular effects on the KATRIN measurement for selected reference values of parameters, showing the sources of systematic error associated with molecular excitations, the projected accuracy on the parameters and the corresponding systematic error on the neutrino mass-squared.   

\begin{table}
\centering
\caption{Summary of molecular-related sources of systematic shift in extracted neutrino mass-squared, the projected accuracy on the experimental parameters and the individual effect on $m_\nu^2$ for the nominal KATRIN parameters shown in Table~\ref{tab:KATRIN-parameters}.  The accuracy of theoretical calculations of the width is taken as 1\% in accordance with the KATRIN Design Report~\cite{katrin04} but further study is necessary to validate this number as discussed in the text.  The achievable experimental uncertainty on the rotational-state temperature is being studied but is not known at this time.}
\label{tab:KATRIN-FSDerrors}
\begin{tabular}{llc}
\hline 
Source of systematic shift & Target accuracy \phantom{aaaaa}& $\sigma_\textrm{syst}(m_\nu^2)[10^{-3}\textrm{eV}^2]$\\

\hline \hline
FSD theoretical calculations \phantom{aa}  & $|\Delta\sigma_\textrm{FSD} / \sigma_\textrm{FSD}| \leq$1\%  & 6 \\ 
temperature calibration 	& $|\Delta T /T| \leq 0.005$ 				& 	\\
\phantom{aaa} - translational	 	&							& 0.05 \\
\phantom{aaa} - FSD 		&								&  0.06\\
temperature fluctuations 	& $|\Delta T /T| \leq 0.001$ 			& 	\\
\phantom{aaa} - translational	 	&							& 0.009 \\
\phantom{aaa} - FSD 		&									& 0.01 \\
ortho-para ratio 			& $|\Delta \lambda / \lambda| \leq 0.03$ 	& 0.44	\\
isotopic impurities			&									&		\\
\phantom{aaa} - tritium purity	&									$|\Delta\epsilon_\textrm{T}/\epsilon_\textrm{T}|\leq0.03 $ 				& 2.9  \\
\phantom{aaa} - ratio of HT to DT 		& $|\Delta \kappa / \kappa| = 0.1$ 		& 0.03 	\\ 
higher rotational states 		& $|\Delta T_{\textrm{rotational}} / T_\textrm{rotational}|$  = 0.1			& 1	\\
\hline
\end{tabular}
\end{table}

\subsection{Summary}
\label{sec:summary}
The use of molecular tritium in experiments to measure the mass of the neutrino necessitates a  quantitative understanding of the role of molecular excitations in modifying the shape of the observed beta spectrum in the vicinity of the endpoint.  Electronic excitations are important but as experimental sensitivity has improved, the focus has increasingly shifted to the rotational and vibrational excitations of the daughter molecule in its electronic ground state.   Those excitations modify the spectrum at the endpoint, whereas the electronic excitations set in some 20 eV below the endpoint.  The KATRIN experiment, by virtue of its high statistical sensitivity and excellent resolution, will be able to concentrate its data-taking in the last 20 eV of the spectrum.

Detailed quantum calculations of the molecular final-state spectrum have been published, and will be used in the analysis of forthcoming experiments.   We have shown that the ground-state rotational and vibrational manifold is fundamentally a Gaussian distribution with a variance determined almost completely by zero-point motion of the nuclei in the parent molecule.   Structure is imposed on that smooth distribution by the quantized nature of the spectrum of final states.  The simplicity of the underlying mechanism suggests that the theoretical prediction of the width of the ground-state manifold should indeed be very reliable, as has been assumed in the design of experiments such as KATRIN.  Calculations using the configuration-interaction method would provide independent uncertainty estimates as well as a comparison to the geminal method calculations.  This would be a significant improvement over the current assessment of errors, which is based solely on the integral of the entire spectrum.  

Thermal excitations of rotational states play a major role for the homonuclear molecule \molT{} since equilibration of the ortho-para ratio is not immediate.  The contribution to the width of the ground-state manifold from rotational-state excitations is relatively small if the molecule is in thermal equilibrium at a temperature near 30 K, but is significant if the distribution remains effectively at 300 K because of the slow thermalization of the ortho-para systems.   Thus the ortho-para ratio must be determined by design or diagnosis.  There is a need for additional theoretical calculations to map out the contributions of states with $J\ge4$. These issues could be circumvented in an experiment that uses HT instead of \molT{}.  Another advantage of using HT is that at 30~K the final-state distribution variance (in the ground-state manifold) is 2/3 as large as it is in \molT.  These advantages are somewhat counterbalanced by a loss of statistical power caused by the dilution of the activity by protium and by the lower source column density caused by the lower molecular mass. 

Although no means is known for a direct experimental measurement of the final-state energy spectrum (other than beta decay itself), the theory makes numerous testable predictions.  The energies of states in the ground-state manifold are in very precise agreement with theory.   Re-evaluating the analysis of the Los Alamos and Livermore gaseous tritium experiments with the current theoretical model produces excellent agreement between the atomic mass difference determined by beta decay and by ion cyclotron resonance.  Furthermore, it eliminates the large negative values of $m_\nu^2$ originally reported in those experiments.   

On the other hand, the measured branching ratios to the bound molecular ions \molion\  and \hhion\   are in the range 90-95\%, in strong disagreement with the theoretical prediction of 39-57\%.  This discrepancy has endured for more than 50 years and a number of possible explanations for it have been suggested.  Several avenues are now open for progress toward a resolution.  New work with the configuration-interaction method is underway~\cite{Saenz:privcomm} and may result in the first independent theoretical cross-check of modern calculations in the geminal basis. A new, direct measurement with beta-ion coincidence information is now feasible with modern instrumentation and is being attempted.  We have presented schematic calculations of the recoil-fragment energy spectra following dissociation, a new and potentially testable aspect of the theory.  Finally, the KATRIN experiment itself will be able to determine the relative fraction of population of the electronic ground and excited states.   With a theoretical cross-check, new experimental information, and insight into the basic mechanism for final-state broadening, one can anticipate increased confidence in quantifying the role it plays when extracting a value for the neutrino mass from data.

\begin{acknowledgments}
The authors wish to thank Alejandro Garc\'{i}a, Alejandro Saenz and Magnus Schl\"{o}sser for many valuable discussions.  This material is based upon work supported by the U.S. Department of Energy Office of Science, Office of Nuclear Physics under Award Number DE-FG02-97ER41020.
\end{acknowledgments}

\bibliography{TRIMS}

\end{document}